\documentclass[11pt,twoside]{book} 
\usepackage{asp2008s}
\usepackage{times}
\usepackage{lscape}
\usepackage{epsf}

\usepackage{graphicx}
\usepackage{amssymb}
\usepackage{longtable}
\usepackage[figuresright]{rotating}
\usepackage{multirow}

\newcommand{\simless}{\mathbin{\lower 3pt\hbox {$\rlap{\raise 5pt\hbox{$\char'074$}}\mathchar"7218$}}}

\mainmatter

\newlength{\deftabcolsep}
\begin{document}

\title{Chamaeleon}   
\author{Kevin L. Luhman}   
\affil{Department of Astronomy and Astrophysics\\
The Pennsylvania State University\\
University Park, PA 16802, USA}

\begin{abstract} 
The dark clouds in the constellation of Chamaeleon have distances
of 160-180~pc from the Sun and a total mass of $\sim$5000~$M_\odot$.
The three main clouds, Cha~I, II, and III, have angular sizes of
a few square degrees and maximum extinctions of $A_V\sim5$-10.
Most of the star formation in these clouds is occurring in Cha~I,
with the remainder in Cha~II.
The current census of Cha~I contains 237 known members, 33 of which
have spectral types indicative of brown dwarfs ($>$M6).
Approximately 50 members of Cha~II have been identified, including a few
brown dwarfs. When interpreted with the evolutionary
models of Chabrier and Baraffe, the H-R diagram for Cha~I exhibits a
median age of $\sim$2~Myr, making it coeval with
IC~348 and slightly older than Taurus ($\sim$1~Myr).
The IMF of Cha~I reaches a maximum at a mass of 0.1-0.15~$M_\odot$, and
thus closely resembles the IMFs in IC~348 and the Orion Nebula Cluster.
The disk fraction in Cha~I is roughly constant at $\sim50$\% from 0.01 to
0.3~$M_\odot$ and increases to $\sim65$\% at higher masses.
In comparison, IC~348 has a similar disk fraction at low masses but a much
lower disk fraction at $M\ga1$~$M_\odot$, indicating that solar-type
stars have longer disk lifetimes in Cha~I.

\end{abstract}

\section{Introduction}

\begin{figure}
\includegraphics[width=\textwidth]{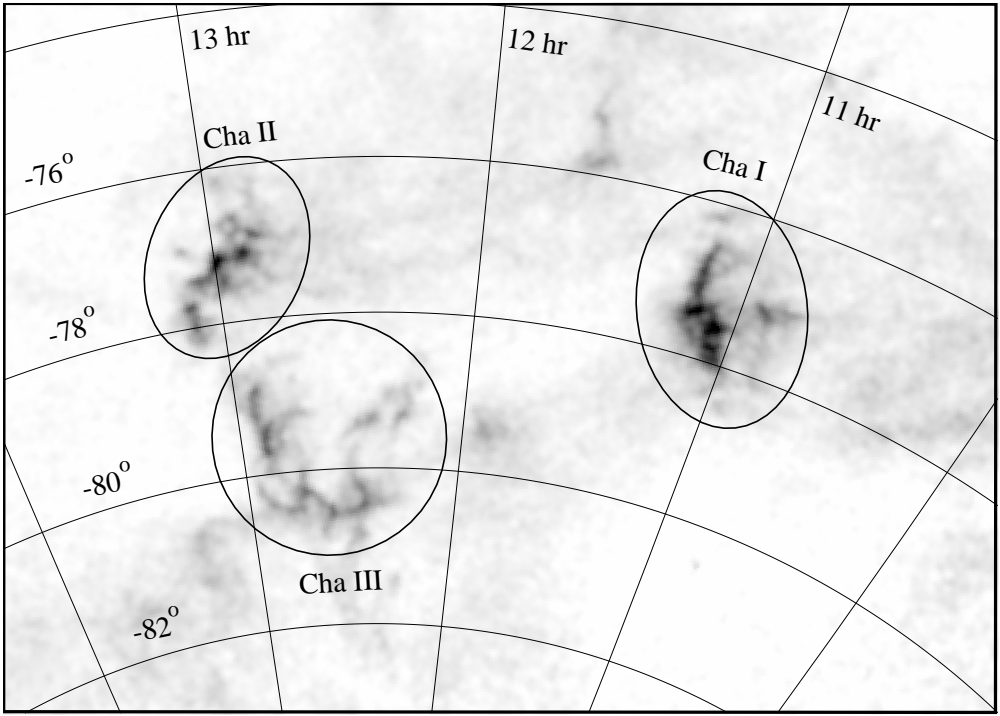}
\caption{Extinction map of the Chamaeleon dark clouds \citep{dob05}.
The maximum extinction in this map is $A_V\sim$10.}
\label{fig:map1}
\end{figure}

\begin{figure}
\includegraphics[width=\textwidth]{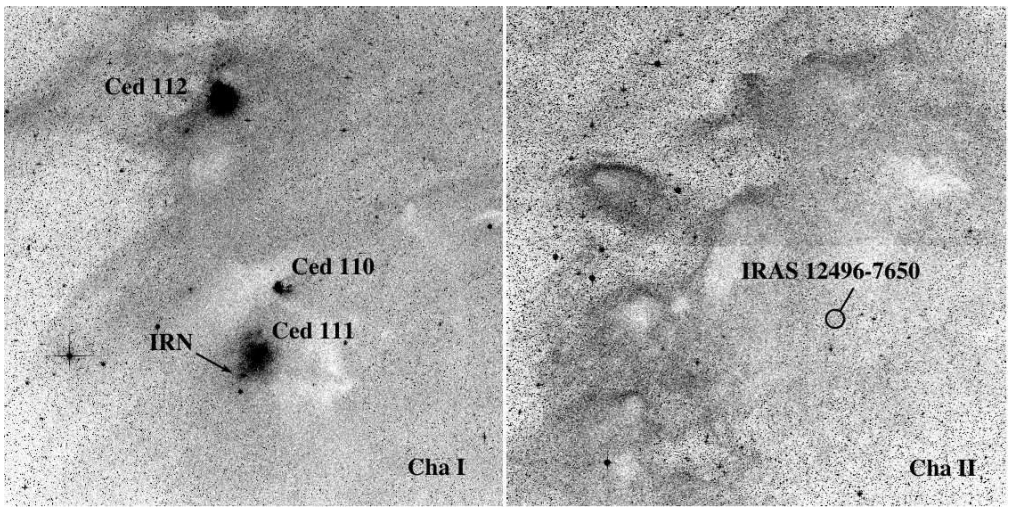}
\caption{
DSS images of Cha~I
($2\hbox{$^\circ$}\times2\hbox{$^\circ$}$) and Cha~II ($1\fdg5\times1\fdg5$).
The reflection nebulae Ced~111 and Ced~112 are associated with the B stars
HD~97048 and HD~97300, respectively.
}
\label{fig:map2}
\end{figure}

\begin{figure}
\includegraphics[width=\textwidth]{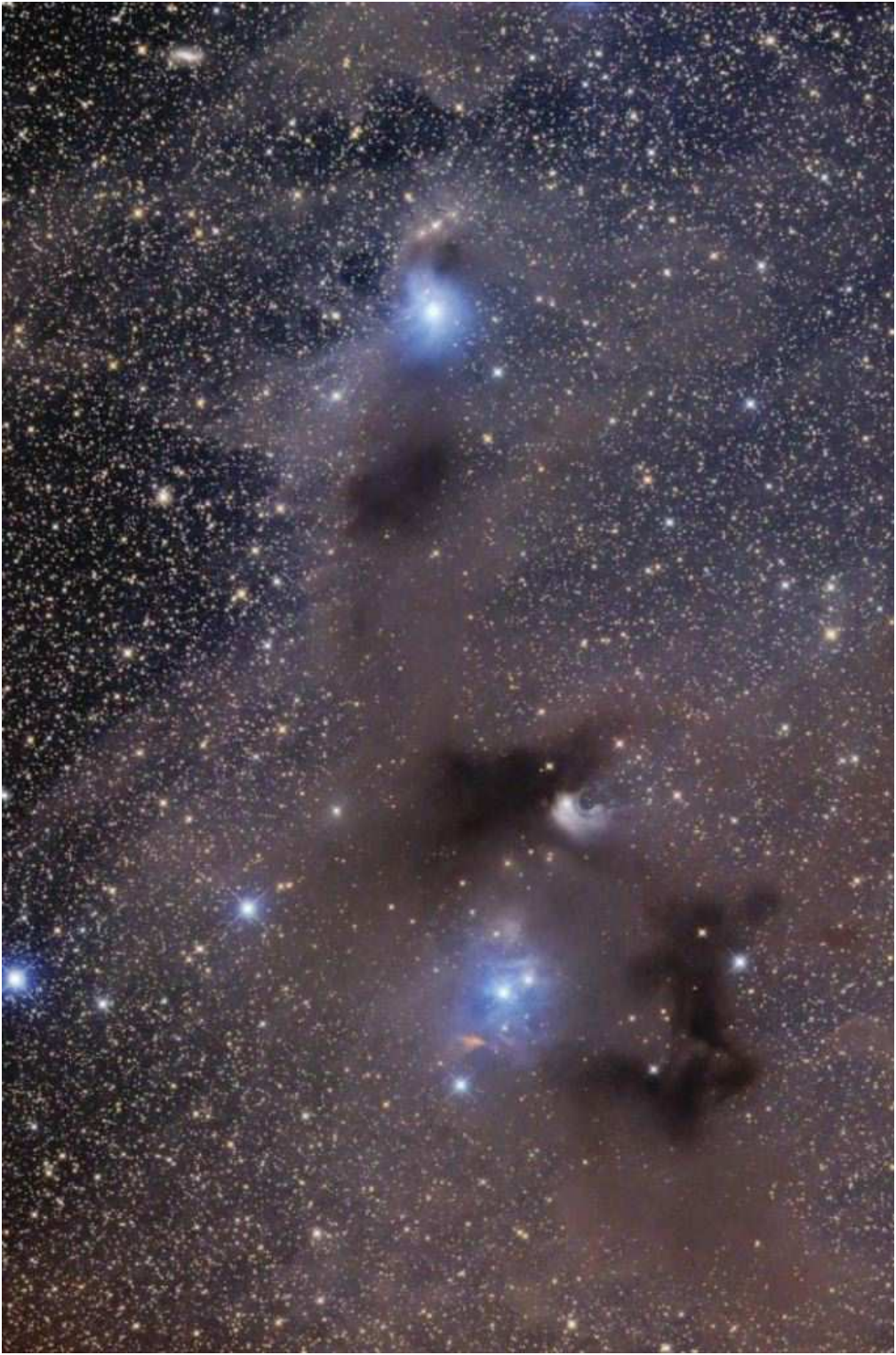}
\caption{
A wide-field optical color-composite image of the Cha~I cloud
($1\fdg4\times2\hbox{$^\circ$}$).
North is up and east is left. Courtesy G. Rhemann.
}
\label{fig:map3}
\end{figure}

\begin{figure}
\includegraphics[scale=1.08]{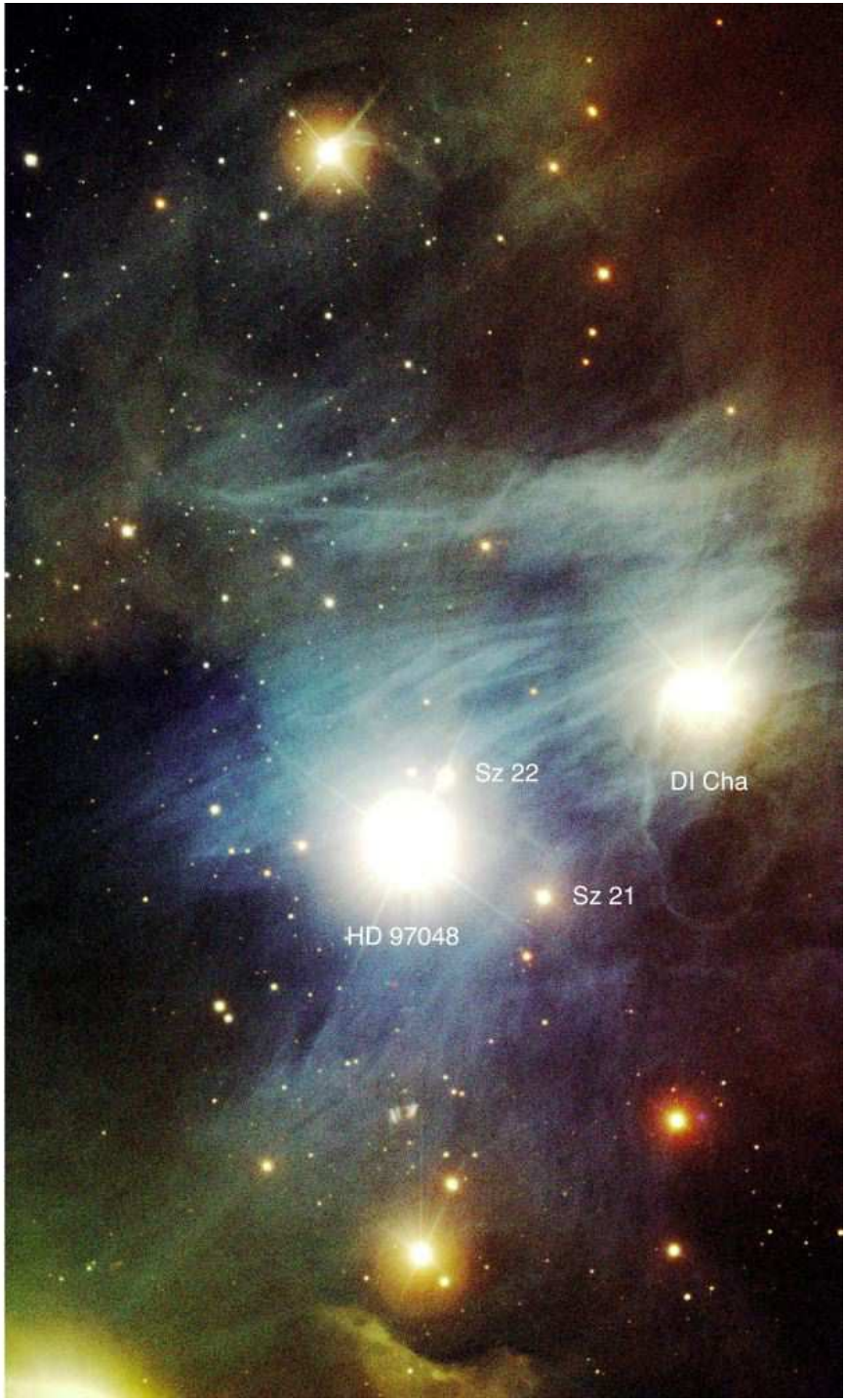}
\caption{
An optical color-composite image of the Ced~111 reflection nebula in Cha~I
obtained with VLT ($6\farcm8\times11\farcm2$).
Four of the brightest young stars within this area are labeled.
North is up and east is left. Courtesy ESO.
}
\label{fig:ced111}
\end{figure}

The southern constellation of Chamaeleon contains one of the nearest
groups of dark clouds to the Sun ($d\sim160$-180~pc). An extinction map of
these clouds is shown in Figure~\ref{fig:map1}
\citep{dob05}. The main clouds in Chamaeleon have angular sizes of
a few square degrees and are referred to as Chamaeleon I, II, and
III \citep{hof62}.\footnote{In most publications, including this review,
the designations Cha~II and Cha~III have been reversed from the original
ones assigned by \citet{hof62}.} Digitized Sky Survey (DSS)
images of Cha~I and II are shown in Figure~\ref{fig:map2} and a color optical
image of Cha~I obtained by G.\ Rhemann is shown in Figure~\ref{fig:map3}.
These clouds contain signposts of recent star formation
in the form of several reflection nebulae, including
Ced~110, 111, \citep{ced46} and the Infrared Nebula
\citep[IRN,][]{sh83}. An optical image of the area surrounding Ced~111
is shown in Figure~\ref{fig:ced111}.
Newborn stars were first directly discovered in
Chamaeleon through their variability and H$\alpha$ emission
\citep{hof62,hen63,men72}.
The masses of the Chamaeleon clouds and the stellar densities of young stars
within them are low compared to many other star-forming regions.
Because Chamaeleon is nearby and well-isolated from other young stellar
populations, it has been a popular target for studies of low-mass star
formation.

\section{Distance}

Published distance estimates for Cha~I have ranged from 115 to 215~pc
\citep{sch91}. The newest measurements by \citet{whi97}, \citet{ber99},
and \citet{wic98} are revisited in this section.
\citet{whi87} measured a distance to Cha~I by examining the distribution
of extinction as a function of distance for stars projected against
the cloud. That analysis was updated with newer photometry by \citet{whi97},
who derived lower and upper limits of 135 and 165~pc. \citet{whi97}
measured a second distance of 152$\pm$18~pc by assuming that
HD~97300, which illuminates Ced~112, was on the zero age main sequence (ZAMS).
They also considered the Hipparcos distances of 190$\pm$40 and 180$\pm$20~pc
for HD~97300 and HD~97048 \citep{per97}. By combining these four distance
constraints, \citet{whi97} arrived at a final value of 160$\pm$15~pc.
Although the strengths of the hydrogen lines in HD~97300 are consistent with
those expected of a ZAMS star \citep{gra75}, the distance based on the ZAMS
assumption is not used in this review.
\citet{ber99} estimated the distance of Cha~I using the Hipparcos measurements
for young stars associated with the cloud. Among those stars, only HD~97300,
HD~97048, and CR~Cha have both definitive evidence of membership in the cloud
and robust Hipparcos distances (i.e., empty H59 field in the Hipparcos
catalog). The weighted average of the parallaxes for these three stars
corresponds to $175^{+20}_{-16}$~pc. In comparison, \citet{wic98}
used the Hipparcos data for CR~Cha, HIP~54738, and T~Cha to estimate
the distance of Cha~I. However, HIP~54738 probably should be omitted because
its H59 field in the Hipparcos catalog indicates that it may be an astrometric
binary and T~Cha is not a member of Cha~I.
Therefore, $175^{+20}_{-16}$~pc appears to be the most appropriate measurement
from Hipparcos for stars in the cloud. The combination of this distance and
the constraint of 135-165~pc from the extinction analysis of \citet{whi97}
indicates a best estimate of 160-165~pc for Cha~I.

The various published distances for Cha~II are summarized by \citet{whi97}.
As they did for Cha~I, \citet{whi97} updated the previous measurements of
extinction as a function of distance for stars toward Cha~II
\citep{fra91,hh92},
arriving at a distance of 178$\pm$18~pc for the cloud. Thus, the distance
of Cha~II is equal to or slightly greater than that of Cha~I.

\section{Cloud Structure}

\begin{figure}[!ht]
\plotfiddle{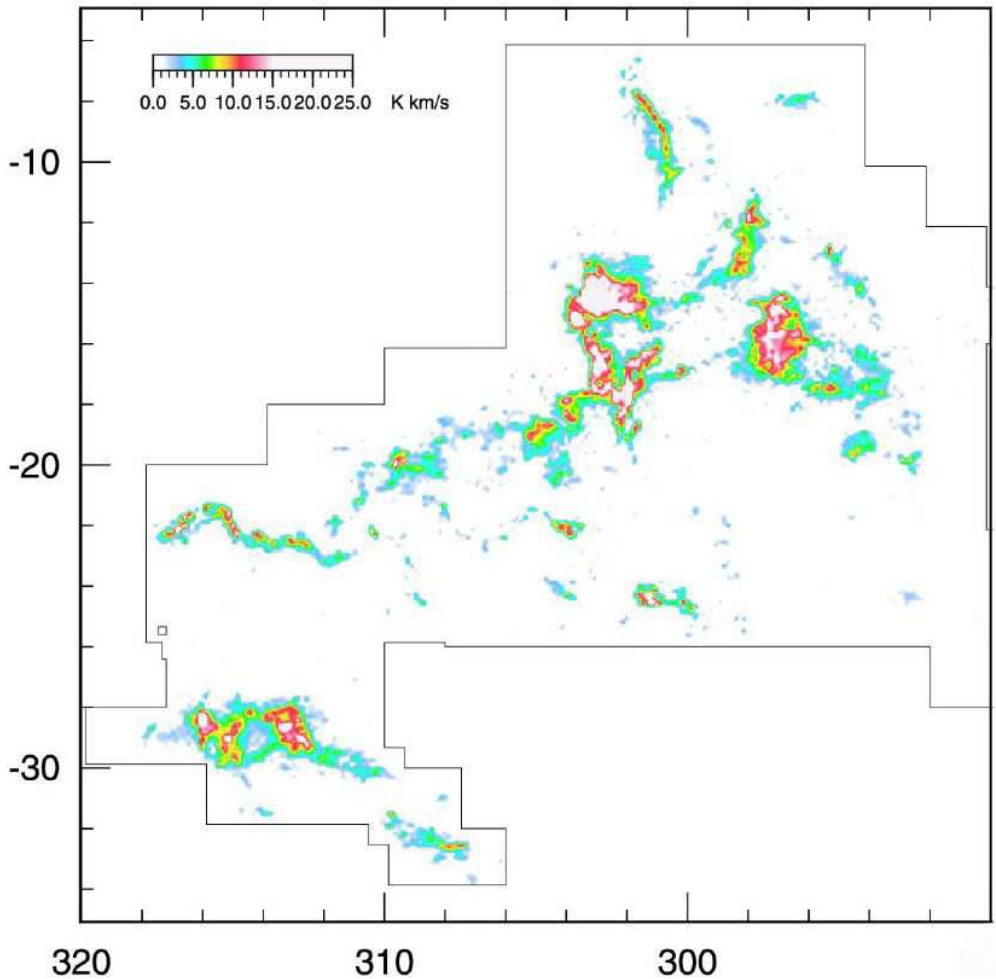}{13.0cm}{0.0}{95.0}{95.0}{-193.0}{0.0}
\caption{
A large-scale $^{12}$CO (J = 1-0) total-intensity map of the
area surrounding the Chamaeleon clouds obtained with the NANTEN
telescope \citep{miz01}. The x-axis is Galactic longitude and the y-axis is
Galactic latitude. The boundaries of the observed area are indicated
({\it solid lines}).
}
\label{fig:mizuno}
\end{figure}

The most obvious signature of a dark cloud is, of course, the darkness
it imposes on the background star population.
Extinction maps with resolutions of a few arcminutes have been measured
for the Chamaeleon clouds using optical \citep{tm85,gre88} and
IR \citep{cam97} star counts. The maximum extinction in these maps is
$A_V\sim$10, which occurs in Cha~I \citep{cam97}. The levels of extinction
in Chamaeleon are modest compared to many other dark clouds.
Chamaeleon also has been mapped in molecules such as
C$^{18}$O \citep{tor90,miz99,hay01,gahm02,hai05},
$^{12}$CO \citep{bou98,miz01}, and $^{13}$CO \citep{miz98,hay01,gahm02}.
The $^{12}$CO map from \citet{miz01} is shown in Figure~\ref{fig:mizuno}.
Based on these maps, Cha~I has a total mass
of $\sim$1000~$M_\odot$ and Cha~II and III each have masses of
1500-2000~$M_\odot$ \citep[e.g.,][]{bou98,miz01}.
Various combinations of maps in molecules, far-IR continuum, H~I,
extinction, and polarization toward field stars have been used to study the
properties of the interstellar medium and magnetic field in the Chamaeleon
clouds \citep{whi94,mcg94,har96,cov97b,bou98,hay99,kai06,nao06}.
A particularly notable characteristic of the interstellar medium in Chamaeleon
is its unusually high value of $R_V\equiv A_V/E(B-V)$, which ranges between 5-6
\citep{gra75,ryd80,vr84,ste89,whi94,whi87,whi97,luh04cha}.
\citet{chi07} obtained mid-IR spectroscopy of field giants behind Cha~I
and other molecular clouds and compared the optical depth of the
9.7~\micron\ silicate feature in these data to near-IR color excesses.
Unlike the diffuse interstellar medium, these clouds did not show
monotonically increasing silicate absorption with extinction, indicating
the presence of grain growth.
Many surveys also have targeted the interstellar medium near individual
dense cores and young stellar objects in Chamaeleon, which have included
maps in CS and CO \citep{olmi94,olmi97,bel06,loh07}, HCN and HNC \citep{ten06},
and continua at far-IR \citep{leh01}, millimeter \citep{hen93,rei96}, and
centimeter \citep{leh03} wavelengths.
One of the most popular targets has been the millimeter source Cha-MMS1
within the Ced~110 reflection nebula, which is probably in the Class~0
stage or in an earlier, prestellar phase \citep{rei96,leh01,leh03,bel06,ten06}.
In addition, the Herbig AeBe stars associated with Ced~111 and 112,
HD~97048 and HD~97300, have been imaged in mid-IR emission bands
\citep{sie98,sie00} and the environment of IRN has been observed
through IR photometry, spectroscopy, and polarimetry
\citep{cs84,fel98,age96,gled96}.
Solid methanol has been detected in the Cha~I cloud through mid-IR spectroscopy
of the Class~I object Cha~INa~2, also known as ISO~192 \citep{pon03}.

\section{Stellar Population}

A variety of signatures of youth have been used to search for young stars
and brown dwarfs in the Chamaeleon clouds, including
H$\alpha$ emission, variability, X-ray emission, and IR excess emission.
Many of the candidates identified with these diagnostics also have been
observed spectroscopically to provide spectral types and conclusive
evidence of youth and membership.
Summaries of these search methods and compilations of known members of
Chamaeleon have appeared periodically in the literature
\citep[e.g.,][]{whi87,gs92,car02,luh04cha}, and updated ones are provided in
this section.

\subsection{H$\alpha$}

Many of the first candidate members of the Chamaeleon clouds were
found during H$\alpha$ objective prism surveys
\citep{hen63,men72,hm73,sch77,har93}.
Followup optical photometry and spectroscopy were then performed on these
candidates \citep{app77,app79,ryd80,app83}.
A deeper H$\alpha$ survey of the southern part of Cha~I
detected 13 members near and below the hydrogen burning mass limit,
which are known as Cha~H$\alpha$~1 through 13 \citep{com99,com00,nc99}.
Since their discovery, these objects have been popular
targets for studies of young low-mass stars and brown dwarfs because of
their relatively close proximity to the Sun and low extinction.
For instance, Cha~H$\alpha$~1 was the first
brown dwarf detected in X-rays \citep{nc98} and several of the Cha~H$\alpha$
sources have been used for studying Li abundances in young low-mass stars
\citep{joh07}. \citet{com04} performed a deep H$\alpha$ survey
across all of Cha~I and obtained spectra of candidates appearing
therein and other candidates from the literature. In doing so, \citet{com04}
identified 18 new members, 15 of which were independently found by
\citet{luh04cha,luh07cha}.

\subsection{Variability}

Along with H$\alpha$ emission, photometric variability was an early indicator
of the presence of young stars in Chamaeleon \citep{hof62}.
The deepest variability survey in Chamaeleon was performed by \citet{car02}.
They conducted near-IR monitoring of a $0\fdg72\times6\hbox{$^\circ$}$ strip
in Cha~I using the southern telescope of the Two-Micron All-Sky Survey
\citep[2MASS,][]{skr06}.
In these data, \citet{car02} discovered 10 objects
that exhibited variability or $K$-band excess emission that were suggestive
of youth and membership in the star-forming region.
Nine of these sources were observed spectroscopically by \citet{luh04cha};
seven candidates were confirmed as members while the others were classified
as background sources.
One of these confirmed members, CHSM~17173, has a spectral type of
\normalsize{M8}, making it the coolest and least massive known member
of Cha~I at that time.

\subsection{X-rays}

\begin{figure}
\centering
\vspace{-8mm}
\includegraphics[scale=0.4,draft=False]{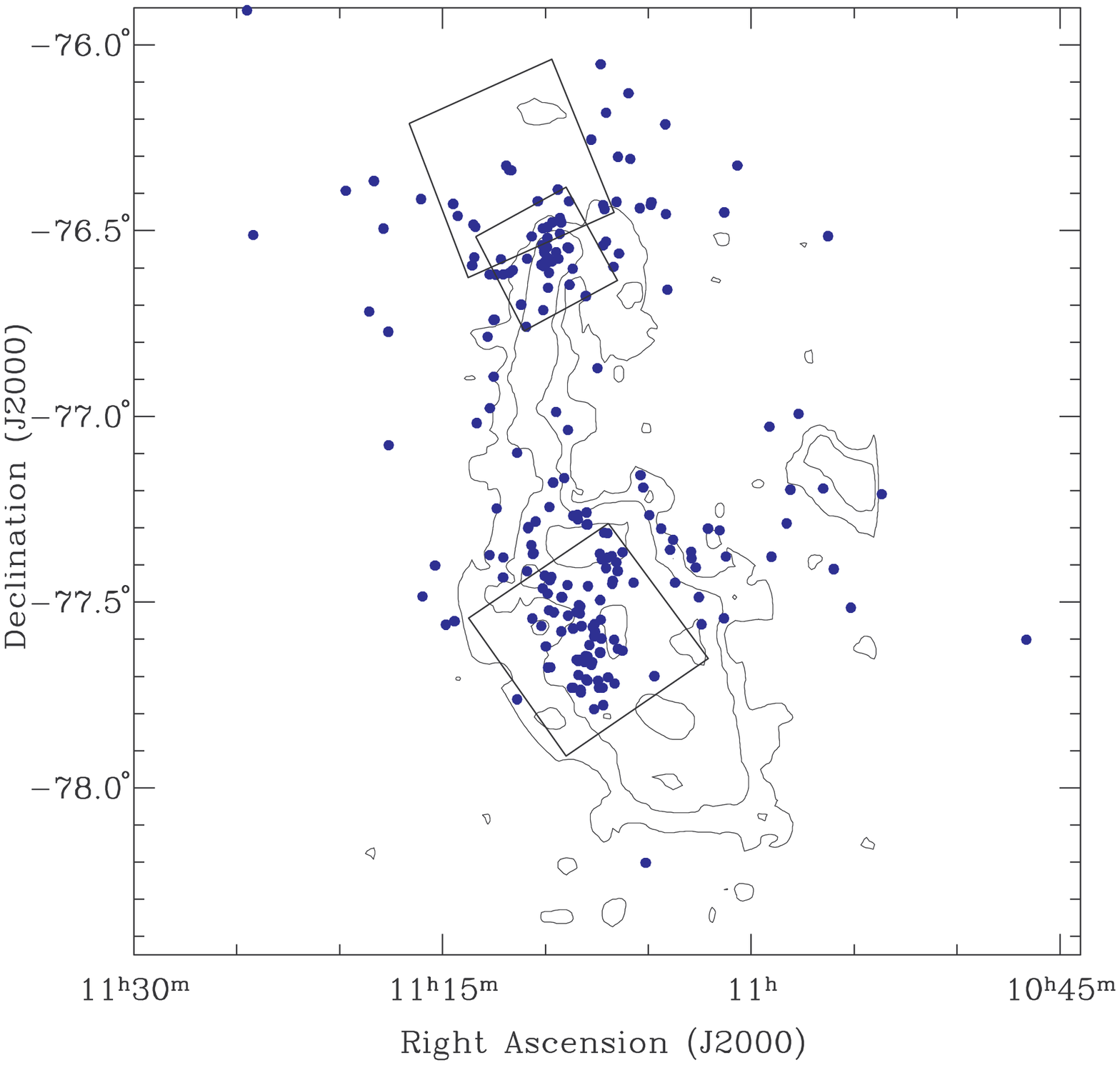}
\caption{
Boundaries of X-ray images from {\it Chandra}
\citep[{\it middle box},][]{fl04} and {\it XMM}
\citep[{\it bottom and top boxes},][]{stel04,rob07} shown with known
members of Cha~I ({\it points}).
The contours represent the extinction map of \citet{cam97} at intervals
of $A_J=$0.5, 1, and 2.
}
\label{fig:xray1}

\includegraphics[scale=0.58,draft=False]{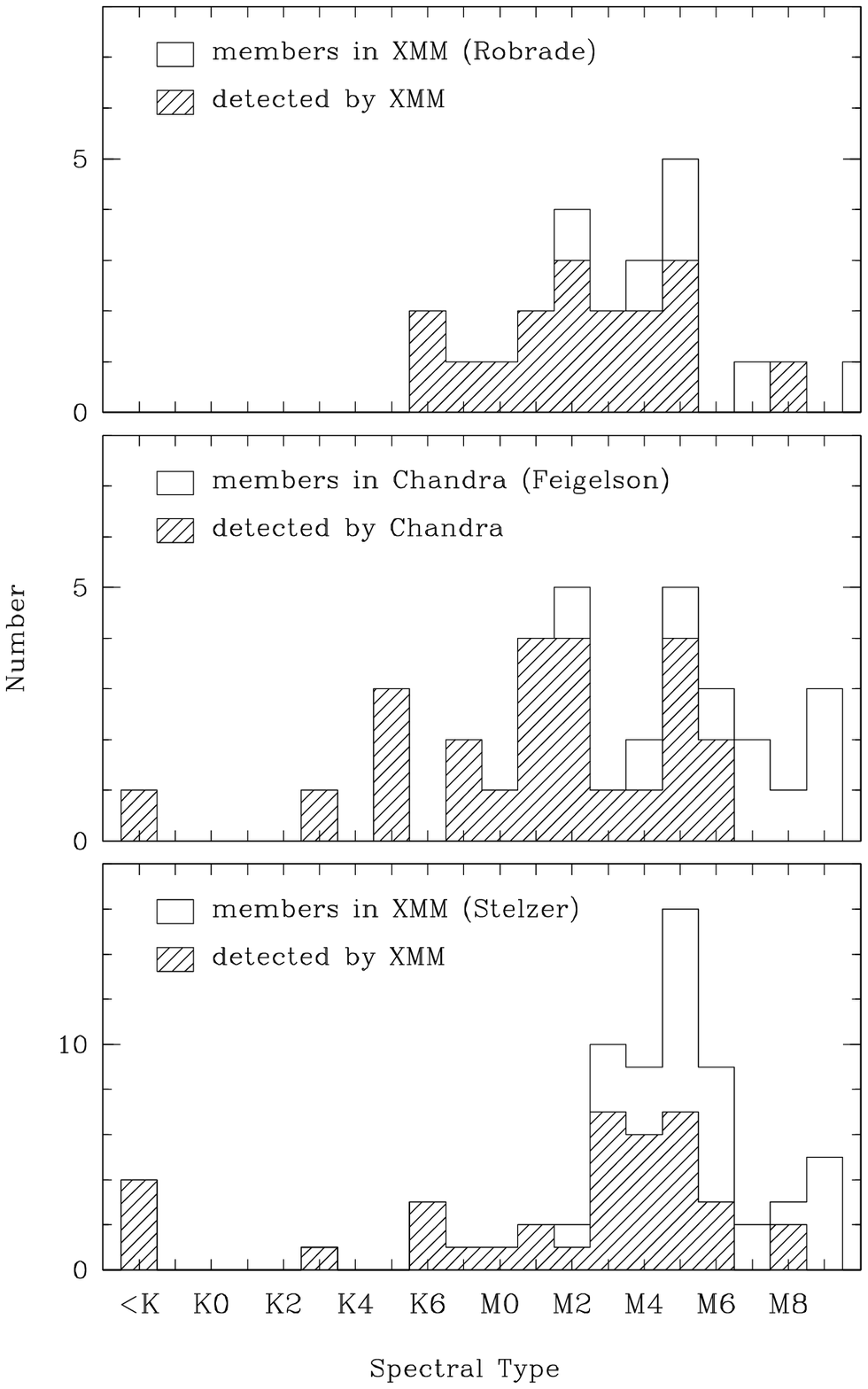}
\caption{
Distributions of spectral types for members of Cha~I within
the X-ray images and the ones that are detected ({\it open and shaded
histograms}).
}
\label{fig:xray2}
\end{figure}

Since the 1980's, X-ray satellites have been heavily employed in searching
for new members of Chamae\-le\-on.
\citet{fk89} obtained X-ray images of most of Cha~I with the
{\it Einstein Observatory} and \citet{wal92} performed spectroscopy on
some of the resulting candidates.
\citet{fei93} detected more candidate members of Cha~I with
the {\it R\"ontgen Satellite} ({\it ROSAT}), which were observed
spectroscopically by \citet{hlf94} and \citet{law96}.
Similar observations have been performed toward Cha~II with {\it ASCA}
\citep{yam98} and {\it ROSAT} \citep{alc00}. \citet{alc95,alc97} and
\citet{cov97a} used the {\it ROSAT} all-sky survey to identify candidate young
stars across the entire Chamaeleon complex and obtained followup optical
spectroscopy for these objects.
The deepest X-ray observations of Chamaeleon have been performed with
{\it XMM-Newton} and the {\it Chandra X-ray Observatory}.
In an exposure with {\it XMM} of
a $27\arcmin\times27\arcmin$ field in the southern cluster in Cha~I,
\citet{stel04} detected 38 of the 72 known members that were within the
field of view.
In the northern cluster in Cha~I, \citet{fl04} observed a
$16\arcmin\times16\arcmin$ field with {\it Chandra}, detecting 27 previously
known members. \citet{fl04} found no new candidate members, and concluded that
the known membership in their survey field is complete to $\sim$0.1~$M_\odot$.
The membership classifications based on the X-ray data from \citet{fl04}
agreed well with the classifications by \citet{luh04cha} using other
diagnostics. \citet{rob07} obtained a deep {\it XMM} image centered on the
young star T49 in Cha~I, detecting 17 of the 23 known members within the field
of view. The boundaries of the {\it Chandra} and {\it XMM} fields and the
distributions of spectral types for all known members of Cha~I within them
are shown in Figures~\ref{fig:xray1} and \ref{fig:xray2}, respectively.
The coolest known members of Cha~I detected in X-rays are
Cha~H$\alpha$~1, Cha~H$\alpha$~7, CHSM~17173, and 2MASS~J11011926$-$7732383B
\citep{nc98,stel04,stel07,rob07}, which have spectral types of M7.75 to \normalsize{M8.25.}

\subsection{Optical and Near Infrared}
\label{sec:optnir}

Because of the modest extinction of most of the Chamaeleon clouds ($A_V<5$),
many of its members can be identified with optical color-magnitude diagrams.
\citet{lm04,lm05} and \citet{spe07} obtained images in H$\alpha$ and broad- and
intermediate-band optical filters for most of the Cha~I and II clouds.
With these data, they identified candidate low-mass members and estimated
their spectral types from the narrowband photometry.
Spectral classifications of the candidates from \citet{lm04} are
summarized by \citet{luh07cha}.
\citet{luh07cha} performed a wide-field search for members of Cha~I using
data from 2MASS and the Deep Near-Infrared Survey of the Southern Sky (DENIS)
and conducted deeper searches in smaller fields using broad-band optical
imaging from Magellan Observatory and the {\it Hubble Space Telescope}.
The candidates from these observations were then observed spectroscopically,
resulting in the confirmation of 50 young stars and brown dwarfs.

Near-IR imaging surveys have been used to search for new members of Chamae\-le\-on
by identifying objects with $K$-band excess emission indicative of
circumstellar disks.
In the earliest study of this kind in Chamaeleon, \citet{hjm82} and
\citet{jon85} performed $K$-band scans of a relatively large area in
Cha~I (1400~arcmin$^2$) and obtained followup $JHK$ photometry and optical
spectroscopy for some of the detected sources to assess their membership.
Using $iJK_s$ photometry from DENIS, \citet{cam98} found 54
candidate members in a $1\fdg5\times3\hbox{$^\circ$}$ field in Cha~I.
Based on spectroscopy and color-magnitude diagrams of these candidates,
\citet{luh04cha} classified 10 as members and 27 as field stars, while
the remaining candidates could not be classified with available data.
Using methods similar to those of \citet{cam98}, \citet{vuo01} produced
a list of 70 candidate low-mass members of Cha~II.
\citet{bar04} performed spectroscopy on 20 of these candidates and confirmed
only one as a member, which they classified as M5.5. This spectral type
corresponds to a mass of $\sim$0.15~$M_\odot$ with the evolutionary models
of \citet{cha00} and \citet{bar98}, placing it near the peak of the stellar
initial mass function (\S~\ref{sec:census}).
By obtaining $JHK$ images of 1~deg$^2$ in Cha~I, \citet{gk01} identified
100 candidate members that exhibited $K$-band excess emission. However, in
the near-IR survey by \citet{car02}, few of these candidates showed excesses.
In an attempt to find members at substellar masses,
\citet{tam98} and \citet{ots99} obtained deeper $JHK$ images of a smaller
field ($5\arcmin\times5\arcmin$) in the northern cluster of Cha~I.
One of the resulting candidate brown dwarfs, OTS~44, has been
spectroscopically confirmed as a member and classified as M9.5
\citep{luh04ots,luh07cha} while another candidate, OTS~32, is a low-mass
star that may have an edge-on disk \citep{luh08cha2}.
In the deepest near-IR observations to date in Cha~I, \citet{cc04} used the
European Southern Observatory's Very Large Telescope to collect $JHK$
images of a $2\farcm5\times2\farcm5$ field. With these near-IR data, they
identified a candidate brown dwarf at $H\sim22$, which would correspond to
a mass below that of Jupiter for a bona fide member. \citet{cc04} obtained
a low-resolution IR spectrum of the candidate, but they could not
conclusively classify the object as either a member or a background source.

\subsection{Mid and Far Infrared}
\label{sec:mirfir}

\begin{figure}[tbp]
\includegraphics[scale=0.7]{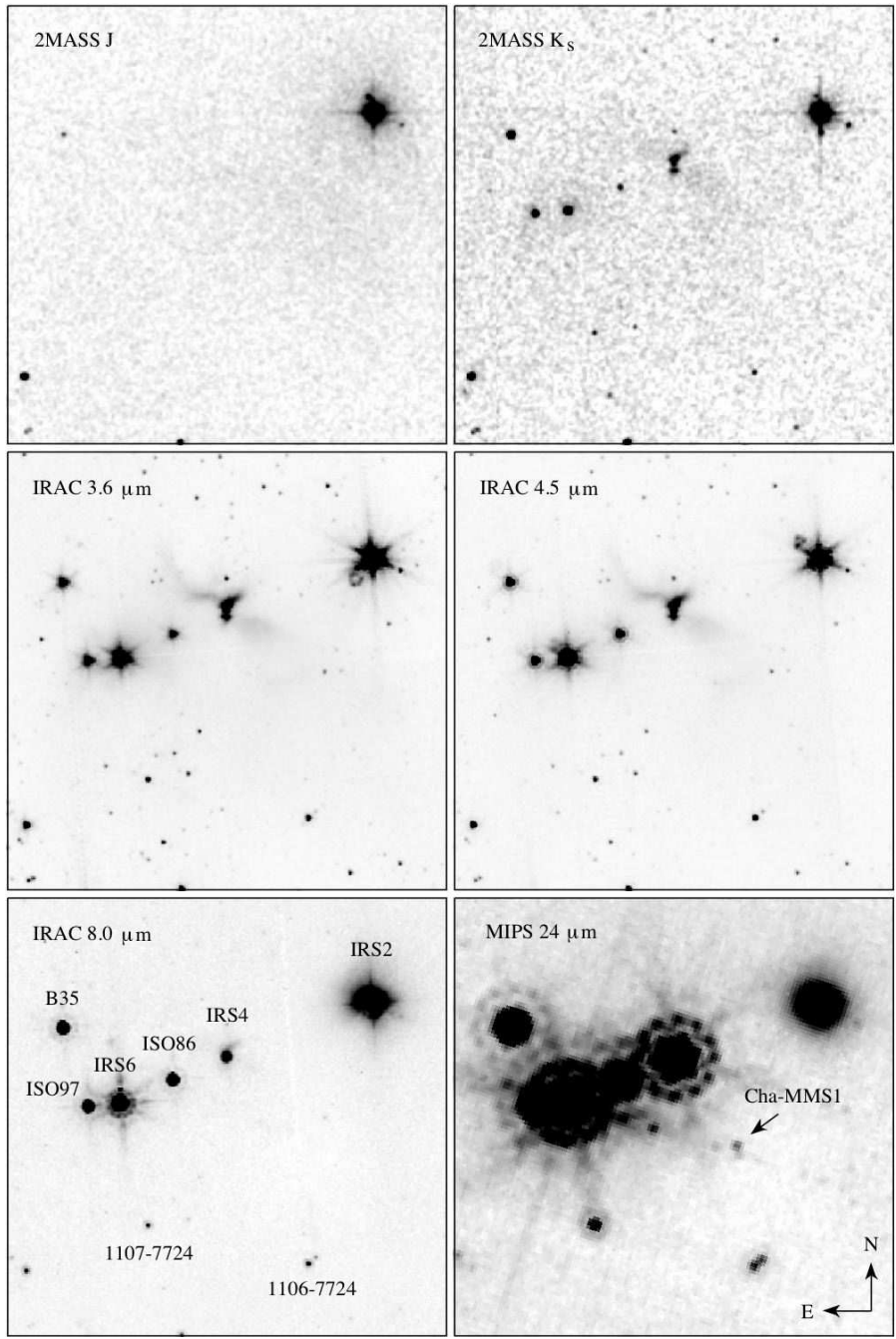}
\caption{
Near- and mid-IR images of the region surrounding the Ced~110 reflection
nebula in Cha~I \citep[$5\arcmin\times5\arcmin$,][]{luh08}.
All known members of Cha~I within this area are labeled.
2MASS J11062942$-$7724586, 2MASS J11070369$-$7724307,
2MASS~1106$-$7724 and 2MASS~1107$-$7724 are highly reddened, low-mass disk-bearing
sources. A redder and fainter candidate member of Cha~I appears
$7\arcsec$ northwest of 2MASS J11062942$-$7724586.
The protostar Cha-MMS1 \citep{rei96} is detected at 24~\micron.
The faint source in the middle of the 24~\micron\ image (southeast of
Ced~110-IRS4) is a latent image from Ced~110-IRS6.
}
\label{fig:image}
\end{figure}

\begin{figure}[!tb]
\centering
\includegraphics[draft=false,width=0.8\textwidth]{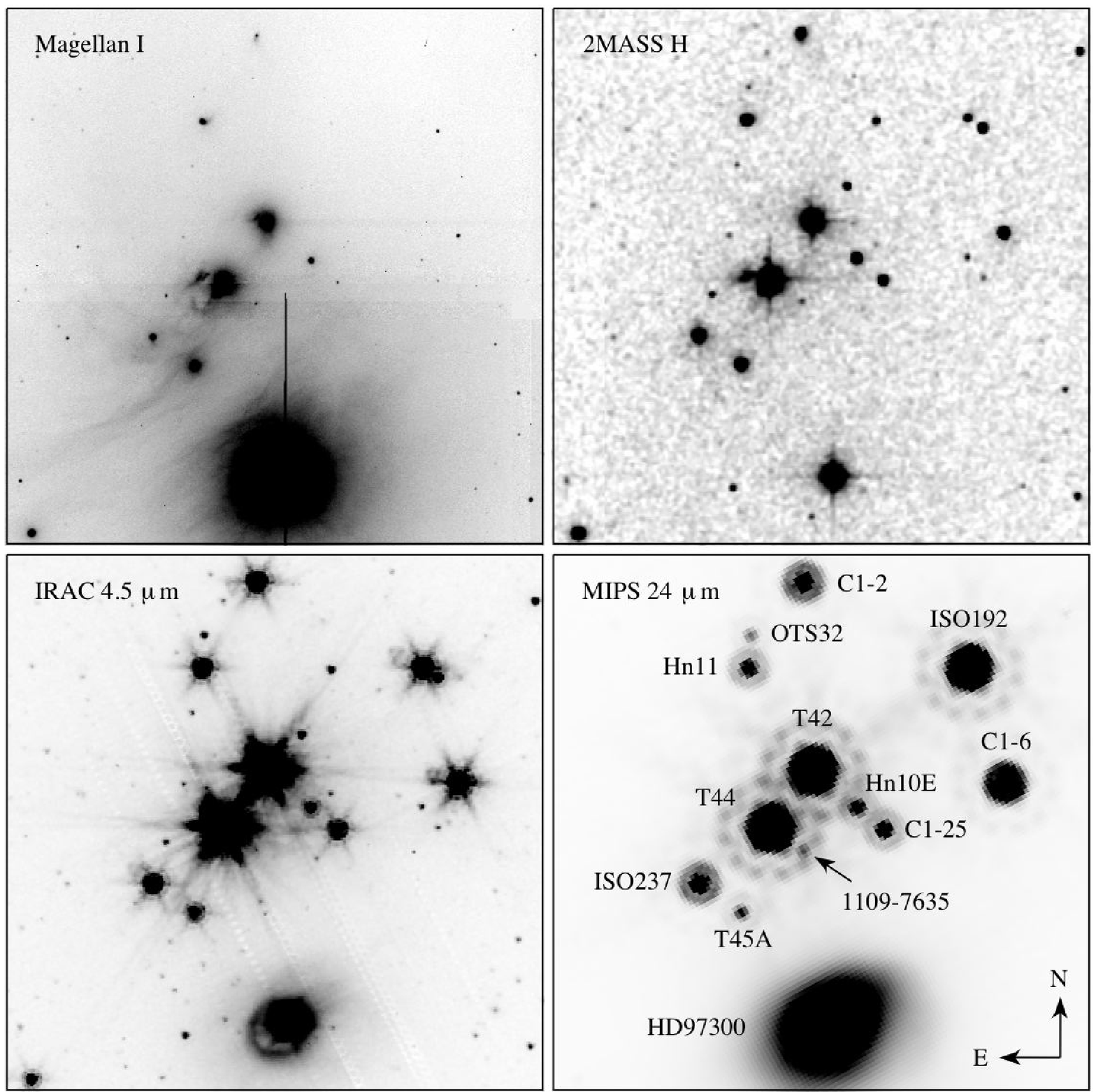}
\caption{
Optical and IR images of the region surrounding the Ced~112 reflection nebula
in Cha~I \citep[$5\arcmin\times5\arcmin$,][]{luh08cha2}.
All known members of Cha~I within this area are labeled.
OTS~32 is a low-mass star that may have an edge-on disk.
2MASS J11095493$-$7635101 is one of the least massive known Class~I sources.
}
\label{fig:image2}
\end{figure}

Compared to near-IR photometry, data at mid- and far-IR wavelengths
are much more sensitive to excess emission from circumstellar material.
Data of this kind are particularly useful for detecting the youngest, most
embedded members of a star-forming region.
Mid- and far-IR searches for new members have been performed with
the {\it Infrared Astronomical Satellite} ({\it IRAS})
for large fractions of Cha~I and II and for
smaller fields around Ced~110, 111, \citep{ass90,pru91}.
Similar surveys were later conducted with the {\it Infrared Space Observatory}
\citep[{\it ISO},][]{nor96,per99,per00,per03,leh01}, which provided greater
sensitivity and higher spatial resolution than {\it IRAS}.
By combining images at 6.7 and 14.3~\micron\ from {\it ISO} with
near-IR data from DENIS, \citet{per00} reported the discovery of 34 new young
stars in Cha~I. Through spectroscopy, \citet{luh04cha} confirmed the membership
of eighteen of these stars and found that 10 others were field stars.
In Cha~II, \citet{alc06} used data from {\it ISO} to identify a candidate
low-mass object, which they spectroscopically confirmed as a member.
The newest IR satellite, the {\it Spitzer Space Telescope},
also has been used to map the Chamaeleon clouds.  \citet{you05} and
\citet{por07} obtained {\it Spitzer} images of Cha~II between 3.6 and
160~$\mu$m. Even with the unprecedented
sensitivity of these data, only a few new young stars were discovered.
Using a combination of $IJHK$ photometry from ground-based telescopes
and mid-IR measurements from {\it Spitzer}, \citet{all06,all07} searched for
young brown dwarfs with disks in Cha~II and other star-forming regions.
Through this work, they recovered the object found by \citet{alc06} with
{\it ISO} and discovered two additional substellar members of Cha~II.
Similarly, \citet{luh05disk2,luh08} and \citet{luh08cha2}
used deep {\it Spitzer} images of Cha~I to find new stars and
brown dwarfs with disks, including the least massive known Class~II and
Class~I members of the cluster and possible edge-on disks.
Some of these new members are within the ground-based and
{\it Spitzer} images of Ced~110 and Ced~112 in
Figures~\ref{fig:image} and \ref{fig:image2}.
The survey of \citet{luh08} demonstrated that the current census of
disk-bearing members of Cha~I is nearly complete at stellar masses except
for close companions and Class~0 sources, while the completeness at
substellar masses is uncertain because of contamination by faint red galaxies.

\vspace{-2mm} 
\subsection{Current Census}
\label{sec:census}

\begin{figure}[tbp]
\includegraphics[draft=False,scale=0.95]{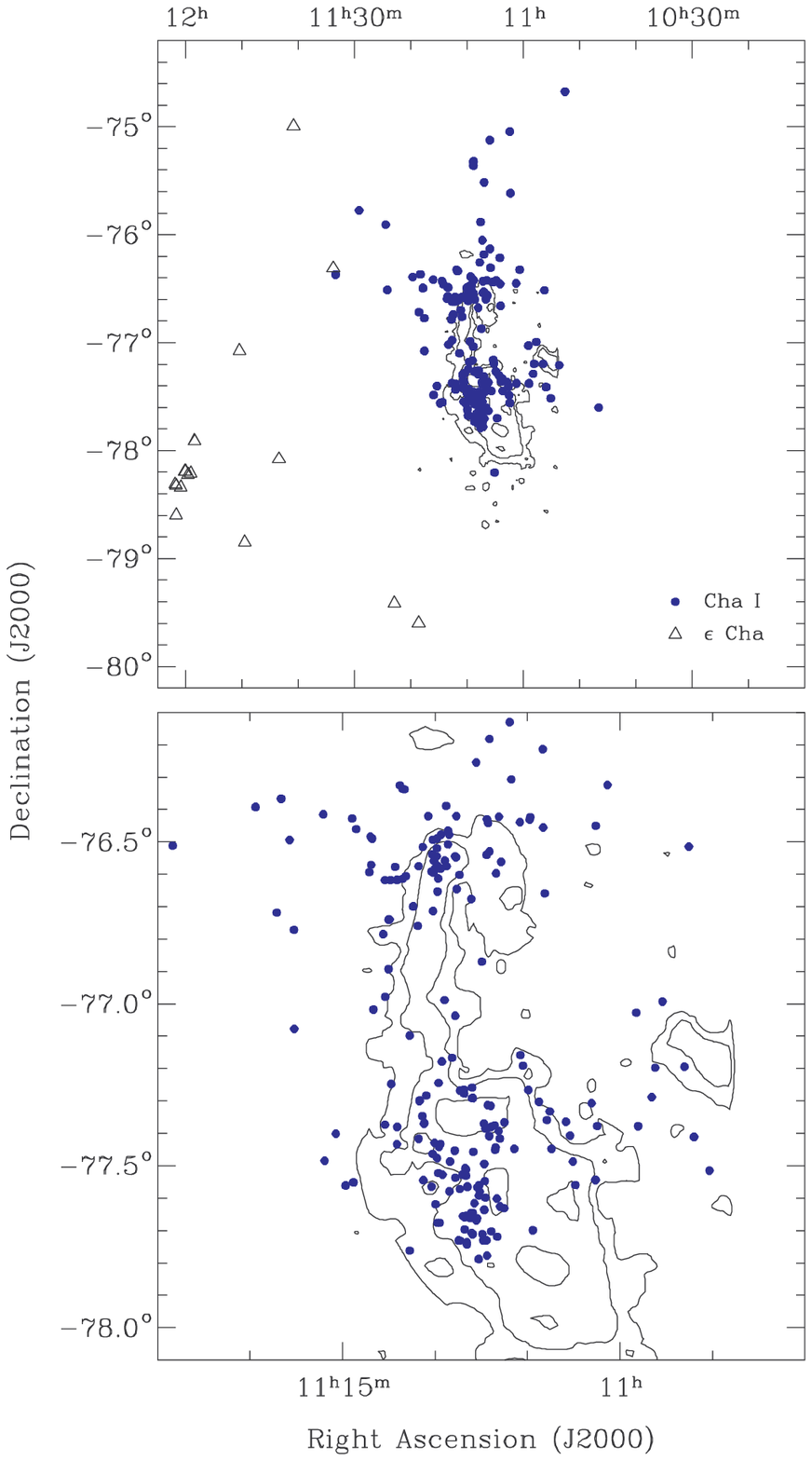}
\caption{
Spatial distribution of known members of Cha~I
\citep[{\it points},][]{luh07cha,luh08}.
Probable members of the $\epsilon$~Cha young association are
indicated \citep[{\it open triangles},][]{cov97a,fei03,luh04eta,luh08}.
The contours represent the extinction map of \citet{cam97} at intervals
of $A_J=$0.5, 1, and 2.
}
\label{fig:map4}
\end{figure}

Although a large number of surveys for young stars have been performed in
Chamae\-le\-on, many of the objects that have been referred
to as members lack either conclusive evidence of membership or accurate
spectral classifications. To refine the census of known members of Cha~I,
\citet{luh04cha} obtained optical spectra of most of the objects that were
previously identified as possible members of the cluster, that lacked either
accurate spectral classifications or evidence of membership, and that were
sufficiently bright
($I\mathrel{\hbox{\rlap{\hbox{\lower4pt\hbox{$\sim$}}}\hbox{$<$}}}$18).
\citet{luh04cha} then used these spectroscopic data and other available
constraints to evaluate the spectral classifications and membership status
of the candidate members of Cha~I that appeared in previous studies of
the cluster. The resulting census contained 158 members.
A likely member that was overlooked in that census is ESO~H$\alpha$~281
from \citet{rz93} and \citet{bz97}.
\citet{luh04cha} assigned membership to SGR~1 and
ISO~13 based on the spectral type from \citet{saf03} and a detection of
mid-IR excess emission from \citet{per00}, respectively. However, they have
been classified as a field star and a galaxy \citep{luh07cha,luh08}.
The census from \citet{luh07cha} did not include protostellar sources
that have been detected only at far-IR wavelengths and longward, such as
Cha-MMS1 \citep{rei96}. Because this source has been detected at
24~\micron\ \citep{bel06,luh08}, it is included in the list of members in
this review. Since the census of \citet{luh04cha}, 78 additional members have
been identified by \citet{com04}, \citet{luh04bin,luh07cha},
\citet{luh04ots,luh05disk2,luh06bin,luh08}, and \citet{luh08cha2}.
\citet{luh07cha} also spectroscopically confirmed the membership of
two candidate companions, T3B and T39B.
Three young stars from \citet{cov97a} share the same proper motion
as Cha~I, and thus are adopted as members of Cha~I \citet{luh08}.
Meanwhile, four young stars from \citet{luh07cha},
2MASS~J11183572$-$7935548, J11334926$-$7618399,
J11404967$-$7459394, and J11432669$-$7804454, are not coun\-ted
as members because their proper motions suggest membership in
$\epsilon$~Cha association rather than the Cha~I cloud \citep{luh08}.
The resulting census of Cha~I contains 237 sources and is compiled in
Tables~1 and 2.
Table~1 lists the various identifications that have been used for each
member, which is an updated version of the compilation of source
names from \citet{car02}. Table~2 provides all available
spectral type measurements and evidence of membership. The latter consists of
$A_V\mathrel{\hbox{\rlap{\hbox{\lower4pt\hbox{$\sim$}}}\hbox{$>$}}}1$
and a position above the main sequence
for the distance of Cha~I ($A_V$), strong emission lines (e),
Na~I and K~I strengths intermediate between those of dwarfs and giants (NaK),
strong Li absorption (Li), IR excess emission (ex),
the shape of the gravity-sensitive steam bands (H$_2$O),
or a proper motion ($\mu$), parallax ($\pi$), or radial velocity
(rv) that is similar to that of the known members of Cha~I
\citep{luh04cha,luh07cha}. The positions of the known members of Cha~I are
plotted with the extinction map from \citet{cam97} in Figure~\ref{fig:map4}.

\begin{figure}[!tb]
\centering
\includegraphics[draft=False,width=0.88\textwidth]{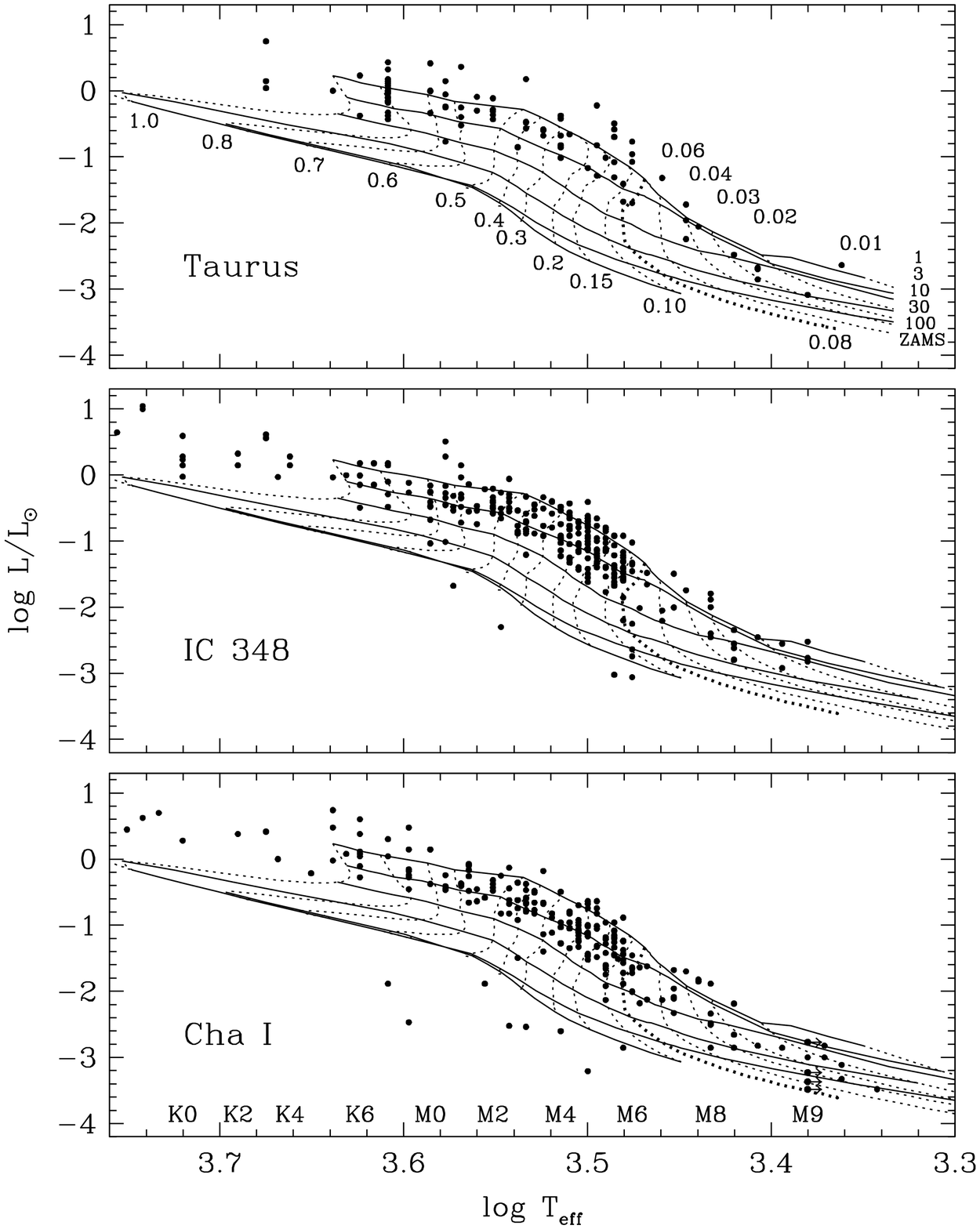}
\caption{H-R diagrams of Cha~I \citep{luh07cha,luh08,luh08cha2},
IC 348 \citep{luh03b}, and Taurus \citep{bri02,luh03a} shown with the
theoretical evolutionary models of \citet[][0.1$<M/M_\odot\leq$1]{bar98} and
\citet[][$M/M_\odot\leq$0.1]{cha00},
where the mass tracks ({\it dotted lines}) and isochrones ({\it solid lines})
are labeled in units of $M_\odot$ and Myr, respectively.}
\label{fig:hr}
\end{figure}

The most recent estimates of extinctions, luminosities, and effective
temperatures for members of Cha~I are provided by \citet{luh07cha},
\citet{luh08}, and \citet{luh08cha2}.
The Hertzsprung-Russell (H-R) diagram for the cluster is shown
in Figure~\ref{fig:hr}. For comparison, H-R diagrams for IC~348 and Taurus
are also included \citep{bri02,luh03a,luh03b}.
According to the evolutionary models of \citet{bar98} and \citet{cha00},
Cha~I has a median age of $\sim$2~Myr, which is similar to that for IC~348
and slightly greater than the age of $\sim$1~Myr for Taurus.
The distribution of isochronal ages suggests that star formation
began 3-4 and 5-6~Myr ago in the southern and northern subclusters,
respectively, and has continued to the present time at a declining rate
\citep{luh07cha}. The combination

\begin{landscape}
\begin{center}
{\footnotesize

\begin{longtable}{cl}
\caption{Known Members of Chamaeleon I: Names}\\

\noalign{\smallskip}
\hline
\noalign{\smallskip}
2MASS & Names \\
\noalign{\smallskip}
\hline
\noalign{\smallskip}
\endfirsthead

\caption{Known Members of Chamaeleon I: Names (continued)}\\
\noalign{\smallskip}
\hline
\noalign{\smallskip}
2MASS & Names \\
\noalign{\smallskip}
\hline
\noalign{\smallskip}
\endhead

\noalign{\smallskip}
\hline
\noalign{\smallskip}
\endfoot

\noalign{\smallskip}
\hline
\noalign{\smallskip}
\endlastfoot

J10463795$-$7736035 &                                                      HD93828 \\
J10523694$-$7440287 &                                                           ~$\cdots$~\\
J10533978$-$7712338 & ~$\cdots$~\\
J10550964$-$7730540 &                                                      ESOH$\alpha$552 \\
 ~$\cdots$~ &                                                          T3B \\
J10555973$-$7724399 &          T3A,Cam1-3,SXCha,Sz2,S6320,HM1,HBC564,B1,10548$-$7708 \\
J10561638$-$7630530 &                                                     ESOH$\alpha$553 \\
J10563044$-$7711393 & T4,Cam1-4,SYCha,Sz3,S6321,HM2,CHXR1,CHX1,HBC565,B2,10552$-$7655 \\
J10574219$-$7659356 &                                                T5,Cam1-5,Sz4 \\
J10580597$-$7711501 &                                                           ~$\cdots$~\\
J10581677$-$7717170 & T6,Cam1-7,SZCha,S6323,CHXR4,CHX2,HBC566,B4,10570$-$7701,GlassV \\
J10590108$-$7722407 &     T7,Cam1-8,TWCha,Sz5,S6326,HM3,CHXR5,HBC567,B5,10577$-$7706 \\
J10590699$-$7701404 & T8,Cam1-9,CRCha,Sz6,HM4,CHXR6,CHX3,HBC244,B6,10578$-$7645,LkH$\alpha$332-20 \\
J11004022$-$7619280 &                                          T10,Cam1-12,Sz8,HM6 \\
J11011370$-$7722387 &                                                           ~$\cdots$~\\
J11011875$-$7627025 &                                               CHXR9C,Cam1-13 \\
J11011926$-$7732383A &                                                           ~$\cdots$~\\
J11011926$-$7732383B &                                                           ~$\cdots$~\\
J11013205$-$7718249 &                                                      ESOH$\alpha$554 \\
J11020610$-$7718079 & ~$\cdots$~\\
J11021927$-$7536576 &                                                           ~$\cdots$~\\
J11022491$-$7733357 & T11,Cam1-15,CSCha,Sz9,HM7,CHXR10,CHX4,HBC569,ISO3,B10,11011$-$7717 \\
J11022610$-$7502407 &                                                           ~$\cdots$~\\
J11023265$-$7729129 &                     CHXR71,Cam1-16,Hn1,ISO4,B11,[LES2004]453 \\
J11024183$-$7724245 &                                                 [LES2004]423 \\
J11025374$-$7722561 & [LES2004]424 \\
J11025504$-$7721508 &                            T12,Cam1-17,Sz10,HM8,[LES2004]456 \\
J11034186$-$7726520 &                                   ISO28,CHSM757,[LES2004]422 \\
J11034764$-$7719563 &                 Hn2,CHSM1012,Cam1-19,ISO32,B13a,[LES2004]425 \\
J11035682$-$7721329 &                        CHXR12,CHSM1403,Cam1-21,Hn3,ISO34,B14 \\
J11040425$-$7639328 &                                             CHSM1715,ESOH$\alpha$557 \\
J11040909$-$7627193 & T14,CHSM1898,Cam1-22,CTCha,Sz11,HM9,CHXR13,HBC570,11027$-$7611 \\
J11041060$-$7612490 &                                                     CHSM1982 \\
J11042275$-$7718080 &                  T14a,CHSM2482,Cam1-24,ISO46,11030$-$7702,HH48 \\
J11044258$-$7741571 &   ISO52,CHSM3319,Cam1-26,B18,[SMN2004]2,ESOH$\alpha$558,[LES2004]457 \\
J11045100$-$7625240 &                                     CHXR14N,CHSM3657,Cam1-27 \\
J11045285$-$7625514 &                                     CHXR14S,CHSM3716,Cam1-28 \\
J11045701$-$7715569 &             T16,CHSM3876,Cam1-29,Sz13,ISO55,B19,[LES2004]603 \\
J11050752$-$7812063 &                                                           ~$\cdots$~\\
J11051467$-$7711290 &                            Hn4,CHSM4629,Cam1-30,[LES2004]622 \\
J11052272$-$7709290 &                                                           ~$\cdots$~\\
J11052472$-$7626209 &                                                           ~$\cdots$~\\
J11054300$-$7726517 &    CHXR15,CHSM5856,Cam1-34,ISO65,B22,[SMN2004]6,[LES2004]449 \\
J11055261$-$7618255 &               T20,CHSM6267,Cam1-35,UVCha,S6336,CHXR18,HBC571 \\
J11055780$-$7607488 &                                                      HD96675 \\
J11060010$-$7507252 &                                                           ~$\cdots$~\\
J11061540$-$7721567 & T21,CHSM7184,Cam1-36,CHXR19,CHX7,ISO75,11048$-$7706,Ced110-IRS2,GlassF,[SMN2004]8 \\
J11061545$-$7737501 &                              Cam2-19,[SMN2004]7,[LES2004]432 \\
J11062554$-$7633418 &                                                      ESOH$\alpha$559 \\
 ~$\cdots$~ &                                                      CHXR73B \\
~$\cdots$~ & ChaJ11062854$-$7618039 \\
J11062877$-$7737331 &           CHXR73A,CHSM7730,ISO78,B25,[SMN2004]9,[LES2004]433 \\
J11062942$-$7724586 &                                                           ~$\cdots$~\\
J11063276$-$7625210 &                                        CHSM7869,[LES2004]712 \\
 ~$\cdots$~ &    Cha-MMS1 \\
J11063799$-$7743090 &                                             ChaH$\alpha$12,CHSM8077 \\
J11063945$-$7736052 &                                          ISO79,CHSM8134,GK13 \\
J11064180$-$7635489 &                            Hn5,CHSM8234,Cam1-37,[LES2004]751 \\
J11064346$-$7726343 & T22,CHSM8284,Cam1-38,UXCha,S6337,ISO80,[SMN2004]10,[LES2004]450 \\
J11064510$-$7727023 & CHXR20,CHSM8369,Cam1-39,CHX8,ISO81,B26,Ced110-IRS3,[SMN2004]11 \\
J11064658$-$7722325 &                Ced110-IRS4,CHSM8415,Cam1-40,ISO84,11051$-$7706 \\
J11065733$-$7742106 &                        CHXR74,CHSM8892,ISO87,B27,[SMN2004]12 \\
J11065803$-$7722488 &                                  ISO86,CHSM8924,Ced110-IRS11 \\
J11065906$-$7718535 & T23,CHSM8978,Cam1-41,UYCha,Sz16,S6340,HM11,ISO89,Ced110-IRS5,[LES2004]452 \\
J11065939$-$7530559 &                                                           ~$\cdots$~\\
J11070324$-$7610565 &                                                           ~$\cdots$~\\
 ~$\cdots$~ &                                                     ESOH$\alpha$281 \\
J11070369$-$7724307 &                                                           ~$\cdots$~\\
 ~$\cdots$~ &                                         ChaJ11070768$-$7626326 \\
J11070919$-$7723049 &                Ced110-IRS6,CHSM9387,Cam1-42,ISO92,11057$-$7706 \\
J11070925$-$7718471 &                                       ISO91,CHSM9391,Cam2-21 \\
J11071148$-$7746394 &        CHXR21,CHSM9467,Cam1-44,Hn6,ISO94,KG1,B28,[SMN2004]13 \\
J11071181$-$7625501 &                                        CHSM9484,[LES2004]711 \\
J11071206$-$7632232 &                 T24,CHSM9496,Cam1-45,UZCha,Sz17,S6341,CHXR75 \\
J11071330$-$7743498 &                 CHXR22E,CHSM9542,Cam1-46,KG2,B29,[SMN2004]15 \\
J11071622$-$7723068 &                                   ISO97,CHSM9652,Cam2-23,KG3 \\
J11071668$-$7735532 &                            ChaH$\alpha$1,CHSM9671,ISO95,[SMN2004]17 \\
J11071860$-$7732516 &                                        ChaH$\alpha$9,CHSM9753,ISO98 \\
J11071915$-$7603048 &             T25,CHSM9782,Cam1-47,Sz18,HM12,HBC572,11057$-$7546 \\
J11072022$-$7738111 &                                                           ~$\cdots$~\\
J11072040$-$7729403 &                               ISO99,[SMN2004]19,[LES2004]448 \\
J11072074$-$7738073 & T26,CHSM9847,Cam1-48,DICha,Sz19,HM13,CHXR23,CHX9,HBC245,ISO100,KG6,B32,11059$-$7721,Ced111-IRS1, \\
 & LkH$\alpha$332-1,[SMN2004]20 \\
J11072142$-$7722117 &                              B35,CHSM9872,Cam2-24,ISO101,KG7 \\
J11072443$-$7743489 &                                                 [LES2004]429 \\
 ~$\cdots$~ &                                         ChaJ11072647$-$7742408 \\
J11072825$-$7652118 & T27,CHSM10138,Cam1-49,VVCha,Sz20,S6342,HM14,CHXR24,HBC573,KG12 \\
J11073519$-$7734493 &         CHXR76,CHSM10394,Cam1-51,ISO106,KG15,B34,[SMN2004]22 \\
J11073686$-$7733335 &         CHXR26,CHSM10464,Cam1-52,ISO108,KG17,B35,[SMN2004]23 \\
J11073775$-$7735308 &                                 ChaH$\alpha$7,CHSM10496,[SMN2004]25 \\
J11073832$-$7747168 &                                          [SMN2004]24,ESOH$\alpha$560 \\
J11073840$-$7552519 &                                                           ~$\cdots$~\\
J11074245$-$7733593 &                     ChaH$\alpha$2,CHSM10685,ISO111,KG20,[SMN2004]26 \\
J11074366$-$7739411 & T28,CHSM10737,Cam1-53,Sz21,HM15,CHXR27,HBC574,ISO112,KG21,B36,[SMN2004]27 \\
J11074610$-$7740089 &                                 ChaH$\alpha$8,CHSM10837,[SMN2004]28 \\
J11074656$-$7615174 &                                                    CHSM10862 \\
J11075225$-$7736569 &                     ChaH$\alpha$3,CHSM11114,ISO116,KG23,[SMN2004]29 \\
J11075588$-$7727257 &  CHXR28,CHSM11256,Cam1-54,CHX10a,ISO117,KG24,B37,[SMN2004]30 \\
J11075730$-$7717262 & CHXR30B,CHSM11315,Cam1-55,KG26,B38b,11065$-$7701,Ced110-IRS7,[LES2004]609 \\
J11075792$-$7738449 &  T29,CHSM11344,Cam1-56,Sz22,HM16,ISO119,KG28,B39,[SMN2004]31 \\
J11075809$-$7742413 &                       T30,CHSM11351,Cam1-57,Sz23,ISO120,KG27 \\
J11075993$-$7715317 &                                         ESOH$\alpha$561,[LES2004]610 \\
J11080002$-$7717304 &      CHXR30A,CHSM11416,Cam1-58,ISO122,KG30,B38a,[LES2004]608 \\
J11080148$-$7742288 & T31,CHSM11462,Cam1-59,VWCha,Sz24,S6343,HM17,CHXR31,CHX11,HBC575,ISO123,KG31,B40, \\
 & Ced111-IRS3,[SMN2004]32 \\
J11080234$-$7640343 &                                                 [LES2004]717 \\
J11080297$-$7738425 &            ISO126,CHSM11529,Cam2-32,KG33,[SMN2004]33,ESOH$\alpha$562 \\
J11080329$-$7739174 & T32,CHSM11547,Cam1-60,CUCha,Sz25,HM18,CHXR29,HBC246,ISO124,KG34,11066$-$7722,Ced111-IRS2, \\
 & HD97048,[SMN2004]34 \\
J11080609$-$7739406 &                                                           ~$\cdots$~\\
 ~$\cdots$~ &                                                         T33B \\
J11081509$-$7733531 & T33A,CHSM12055,Cam1-61,CHXR32,CHX12,ISO135,KG39,B42,11068$-$7717,Ced111-IRS4,GlassI,[SMN2004]37 \\
J11081648$-$7744371 &  T34,CHSM12090,Cam1-62,Sz26,HM19,ISO136,KG40,B44,[SMN2004]38 \\
J11081703$-$7744118 &                    ChaH$\alpha$13,CHSM12108,ISO137,KG41,[SMN2004]39 \\
J11081850$-$7730408 &                                ISO138,CHSM12159,[LES2004]410 \\
J11081896$-$7739170 &                                 ChaH$\alpha$4,CHSM12180,[SMN2004]40 \\
 ~$\cdots$~ &                                         ChaJ11081938$-$7731522 \\
J11082238$-$7730277 &                           ISO143,CHSM12355,KG45,[LES2004]411 \\
J11082404$-$7739299 &                                            ChaH$\alpha$10,CHSM12438 \\
J11082410$-$7741473 &                     ChaH$\alpha$5,CHSM12442,ISO144,KG47,[SMN2004]41 \\
J11082570$-$7716396 &                                                           ~$\cdots$~\\
J11082650$-$7715550 &                                ISO147,CHSM12525,[LES2004]613 \\
J11082927$-$7739198 &                                            ChaH$\alpha$11,CHSM12653 \\
 ~$\cdots$~ &                                         ChaJ11083040$-$7731387 \\
J11083896$-$7743513 &     IRN,CHSM13053,Cam1-63,ISO150,KG49,11072$-$7727,Ced111-IRS5 \\
J11083905$-$7716042 & T35,CHSM13058,Cam1-64,Sz27,HM20,C6-1,HBC576,ISO151,KG50,Ced110-IRS8 \\
J11083952$-$7734166 &                     ChaH$\alpha$6,CHSM13082,ISO152,KG51,[SMN2004]43 \\
J11084069$-$7636078 & CHXR33,CHSM13131,Cam1-65,CHX13a,C1-10,ISO153,KG52,[FL2004]8,[LES2004]752 \\
J11084296$-$7743500 &                                                           ~$\cdots$~\\
J11084952$-$7638443 &                                                           ~$\cdots$~\\
J11085090$-$7625135 & T37,CHSM13553,Cam1-68,Sz28,ISO157,KG54,[FL2004]15,[LES2004]709 \\
J11085176$-$7632502 &                                                 [LES2004]737 \\
J11085242$-$7519027 &                                              RX J1108.8$-$7519B \\
J11085326$-$7519374 &                                              RX J1108.8$-$7519A \\
J11085367$-$7521359 & ~$\cdots$~\\
J11085421$-$7732115 &        CHXR78C,CHSM13696,Cam1-69,ISO160,KG56,B45,[SMN2004]45 \\
J11085464$-$7702129 & T38,CHSM13707,Cam1-70,VYCha,Sz29,S6346,HM21,C4-5,HBC577,ISO162,KG58 \\
J11085497$-$7632410 &                   ISO165,CHSM13720,KG59,ESOH$\alpha$564,[LES2004]735 \\
J11085596$-$7727132 &                                                       ISO167 \\
J11090512$-$7709580 &                    Hn7,CHSM14124,Cam1-71,ISO174,[LES2004]615 \\
J11091172$-$7729124 & T39A,CHSM14388,Cam1-72,Sz30,CHXR36,CHX14,C7$-$7,ISO182,KG74,B46,[SMN2004]47 \\
 ~$\cdots$~ &                                                         T39B \\
J11091297$-$7729115 & ~$\cdots$~\\
 ~$\cdots$~ &                                         ChaJ11091363$-$7734446 \\
J11091380$-$7628396 & CHXR35,CHSM14469,Cam1-73,Hn8,C1-15,KG75,[FL2004]25,[LES2004]708,[RS2007]4 \\
J11091769$-$7627578 &    CHXR37,CHSM14604,Cam1-74,ISO185,KG77,[FL2004]26,[RS2007]5 \\
J11091812$-$7630292 & CHXR79,CHSM14623,Cam1-75,Hn9,C1-18,ISO186,KG79,[FL2004]27,[RS2007]6 \\
J11092266$-$7634320 & C1-6,CHSM14800,Cam1-76,ISO189,KG82,OTS10,Ced112-IRS2,[FL2004]29,[LES2004]731 \\
J11092379$-$7623207 & T40,CHSM14843,Cam1-77,VZCha,Sz31,S6347,HM22,CHXR39,HBC578,KG83,11078$-$7607,Ced112-IRS1,[RS2007]7 \\
J11092855$-$7633281 &               ISO192,CHSM15044,Cam2-41,KG87,OTS15,[FL2004]33 \\
J11092913$-$7659180 &                                                 [LES2004]602 \\
J11093543$-$7731390 &                                                           ~$\cdots$~\\
J11093777$-$7710410 &                                                      Cam2-42 \\
J11094006$-$7628391 & CHXR40,CHSM15504,Cam1-78,CHX15b,ISO198,KG92,[FL2004]38,[RS2007]9 \\
J11094192$-$7634584 &         C1-25,CHSM15576,Cam1-79,ISO199,KG93,OTS19,[FL2004]41 \\
J11094260$-$7725578 & C7-1,CHSM15600,Cam1-80,ISO200,KG95,Ced110-IRS9,ESOH$\alpha$565,[LES2004]418 \\
J11094525$-$7740332 &                             [SMN2004]51,ESOH$\alpha$566,[LES2004]434 \\
J11094621$-$7634463 & Hn10E,CHSM15745,Cam1-82,T42,Sz32,HM23,C1-24,ISO204,KG97,OTS24,11083$-$7618,Ced112-IRS4,[FL2004]43 \\
J11094742$-$7726290 & B43,CHSM15784,Cam1-84,ISO207,KG101,[SMN2004]52,ESOH$\alpha$567,[LES2004]417 \\
J11094866$-$7714383 &                                     ISO209,CHSM15850,Cam2-45 \\
J11094918$-$7731197 &                                                        KG102 \\
J11095003$-$7636476 & T41,CHSM15900,Cam1-85,CHXR42,C1-11,ISO211,KG103,OTS26,11082$-$7620,Ced112-IRS3,HD97300,[FL2004]44 \\
J11095215$-$7639128 &          ISO217,CHSM15980,KG106,GK29,[FL2004]46,[LES2004]726 \\
J11095262$-$7740348 &                                                    CHSM15991 \\
J11095336$-$7728365 &                           ISO220,CHSM16023,GK30,[LES2004]412 \\
J11095340$-$7634255 & T42,CHSM16026,Cam1-86,Sz32,HM23,C1-5,HBC579,ISO223,KG109,OTS27,11083$-$7618,[FL2004]47,[LES2004]753 \\
J11095407$-$7629253 & T43,CHSM16052,Cam1-87,Sz33,CHXR41,CHX15,C1-17,ISO224,KG110,[FL2004]48,[RS2007]12 \\
J11095437$-$7631113 &                             ISO225,CHSM16067,GK31,[FL2004]49 \\
J11095493$-$7635101 & ~$\cdots$~\\
J11095505$-$7632409 & C1-2,CHSM16096,Cam1-88,ISO226,KG111,OTS28,Ced112-IRS5,[FL2004]51 \\
J11095873$-$7737088 & T45,CHSM16254,Cam1-89,WXCha,Sz35,S6349,HM25,CHXR43,CHX17,C8-3,HBC581,ISO228,KG114,B48, \\
 & 11085$-$7720,Ced111-IRS6,[SMN2004]54 \\
J11100010$-$7634578 & T44,CHSM16307,Cam1-90,WWCha,Sz34,S6348,HM24,CHXR44,C1-7,HBC580,ISO231,KG116,OTS29,[FL2004]53 \\
J11100192$-$7725451 &                                                 [LES2004]419 \\
J11100336$-$7633111 & OTS32 \\
J11100369$-$7633291 &    Hn11,CHSM16436,Cam1-91,C1-4,ISO232,KG117,OTS33,[FL2004]55 \\
J11100469$-$7635452 & T45a,CHSM16470,Cam1-92,CHXR45,CHX16,C1-9,HBC582,ISO233,KG118,OTS35,GK-1,[FL2004]56 \\
J11100658$-$7642486 &                                                           ~$\cdots$~\\
J11100704$-$7629376 & T46,CHSM16560,Cam1-93,WYCha,Sz36,S6351,HM26,CHXR46,C1-16,HBC583,ISO234,KG119,11085$-$7613, \\
 & Ced112-IRS6,[FL2004]58,[RS2007]13 \\
J11100785$-$7727480 &                                       ISO235,CHSM16611,KG120 \\
J11100934$-$7632178 &                                              OTS44,CHSM16658 \\
J11101141$-$7635292 & ISO237,CHSM16751,Cam2-48,C1-8,KG121,OTS45,[FL2004]61,[LES2004]760 \\
J11101153$-$7733521 &                                                           ~$\cdots$~\\
J11102226$-$7625138 &                            CHSM17173,[LES2004]710,[RS2007]16 \\
J11102852$-$7716596 &             Hn12W,CHSM17417,Cam1-94,C6-2,ISO244,[LES2004]625 \\
J11103481$-$7722053 &                                                 [LES2004]405 \\
J11103644$-$7722131 &                                ISO250,CHSM17743,[LES2004]404 \\
J11103801$-$7732399 & CHXR47,CHSM17807,Cam1-96,F34,C7-11,ISO251,KG124,B50,11091$-$7716,Ced111-IRS7,GlassQ,[SMN2004]55 \\
J11104006$-$7630547 &                                                           ~$\cdots$~\\
J11104141$-$7720480 &                                ISO252,CHSM17959,[LES2004]406 \\
J11104959$-$7717517 &     T47,CHSM18261,Cam1-97,Sz37,HM27,HBC584,ISO254,11093$-$7701 \\
J11105076$-$7718031 &                                         ESOH$\alpha$568,[LES2004]408 \\
J11105333$-$7634319 & T48,CHSM18403,Cam1-98,WZCha,Sz38,S6352,HM28,CHXR82,C1-23,HBC585,ISO258,[FL2004]74 \\
J11105359$-$7725004 &                        ISO256,CHSM18416,Cam2-49,[LES2004]403 \\
J11105597$-$7645325 &   Hn13,CHSM18502,Cam1-99,C2-5,ISO259,[FL2004]75,[LES2004]755 \\
J11111083$-$7641574 &                                                      ESOH$\alpha$569 \\
J11112249$-$7745427 &                                                 [LES2004]436 \\
J11112260$-$7705538 &                                          ISO274,[LES2004]629 \\
J11113474$-$7636211 &             CHXR48,CHSM20217,Cam1-101,E1-7,ISO280,[FL2004]91 \\
J11113965$-$7620152 & T49,CHSM20409,Cam1-102,XXCha,Sz39,HM29,HBC586,11101$-$7603,[RS2007]35 \\
J11114533$-$7636505 &                                                 [LES2004]743 \\
J11114632$-$7620092 &              CHX18N,CHSM20702,Cam1-103,11101$-$7603,[RS2007]37 \\
J11115400$-$7619311 &            CHXR49NE,CHSM21078,Cam1-104,CHX18,Hn15,[RS2007]39 \\
J11120288$-$7722483 &                                                           ~$\cdots$~\\
J11120327$-$7637034 & CHXR84,CHSM21466,Cam1-105,Hn16,E1-10,[FL2004]100,[LES2004]742 \\
J11120351$-$7726009 &                        ISO282,CHSM21473,ESOH$\alpha$570,[LES2004]438 \\
J11120984$-$7634366 & T50,CHSM21745,Cam1-106,Sz40,CHXR85,E1-5,HBC587,[FL2004]103,[LES2004]757,[RS2007]44 \\
J11122250$-$7714512 &                                                           ~$\cdots$~\\
J11122441$-$7637064 & T51,CHSM22361,Cam1-107,Sz41,CHXR50,CHX20b,E1-9a,HBC588,11108$-$7620,Ced112-IRS7,[FL2004]105 \\
J11122772$-$7644223 & T52,CHSM22521,Cam1-108,CVCha,Sz42,HM30,CHXR51,CHX19,E2-4,HBC247,11108$-$7627,Ced112-IRS8,LkH$\alpha$332-21 \\
J11123092$-$7644241 &                T53,CHSM22662,Cam1-110,CWCha,Sz43,HM31,HBC589 \\
J11123099$-$7653342 &                                                 [LES2004]601 \\
J11124210$-$7658400 &                        CHXR54,CHSM23146,Cam1-111,CHX21a,E4-3 \\
J11124268$-$7722230 &    T54,CHSM23172,Cam1-112,HMAnon,CHXR56,CHX22,B51,11111$-$7705 \\
J11124299$-$7637049 &                        CHXR55,CHSM23183,Cam1-113,CHX20E,E1-8 \\
J11124861$-$7647066 &                    Hn17,CHSM23447,Cam1-114,E2-9,[LES2004]754 \\
J11132012$-$7701044 &                           CHXR57,CHSM24835,Cam1-115,E4-4,B52 \\
J11132446$-$7629227 &              Hn18,CHSM25026,Cam1-116,[LES2004]747,[RS2007]55 \\
J11132737$-$7634165 &                    CHXR59,CHSM25148,Cam1-117,E1-4,[RS2007]58 \\
J11132970$-$7629012 &       CHXR60,CHSM25253,Cam1-118,Hn19,[LES2004]748,[RS2007]60 \\
J11133356$-$7635374 &         T55,CHSM25420,Cam1-119,Sz44,CHXR61,E1-6,[LES2004]759 \\
J11141565$-$7627364 &                    CHXR62,CHSM27295,Cam1-120,Hn20,[RS2007]65 \\
J11142454$-$7733062 &                                     Hn21W,CHSM27714,Cam1-121 \\
J11142611$-$7733042 &                                     Hn21E,CHSM27789,Cam1-122 \\
J11142906$-$7625399 &                                           ESOH$\alpha$571,[RS2007]67 \\
J11145031$-$7733390 &                                       B53,CHSM28885,Cam1-123 \\
J11152180$-$7724042 &                                                      ESOH$\alpha$572 \\
J11155827$-$7729046 &                                                           ~$\cdots$~\\
J11160287$-$7624533 &                                                      ESOH$\alpha$574 \\
J11173700$-$7704381 &              T56,Cam1-125,Sz45,HM32,CHXR66,HBC590,11159$-$7648 \\
J11173792$-$7646193 &                                                           ~$\cdots$~\\
J11175211$-$7629392 &                                                           ~$\cdots$~\\
J11181957$-$7622013 &                                                      CHXR68B \\
J11182024$-$7621576 &                                             CHXR68A,Cam1-126 \\
J11183379$-$7643041 &                                                           ~$\cdots$~\\
J11194214$-$7623326 &                                                           ~$\cdots$~\\
J11195652$-$7504529 &                                                           ~$\cdots$~\\
J11241186$-$7630425 &                                                           ~$\cdots$~\\
J11242980$-$7554237 &                                                           ~$\cdots$~\\
J11291261$-$7546263 &                                               RX J1129.2$-$7546 \\
J11332327$-$7622092 &                                                           ~$\cdots$~\\

\end{longtable}


\begin{longtable}{ccccc}
\caption{Known Members of Chamaeleon I: Spectral Types and Membership}\\

\noalign{\smallskip}
\hline
\noalign{\smallskip}
Name & Spectral Type & References & Membership & References \\
\noalign{\smallskip}
\hline
\noalign{\smallskip}
\endfirsthead

\caption{Known Members of Chamaeleon I: Spectral Types and Membership (continued)}\\
\noalign{\smallskip}
\hline
\noalign{\smallskip}
Name & Spectral Type & References & Membership & References \\
\noalign{\smallskip}
\hline
\noalign{\smallskip}
\endhead

\noalign{\smallskip}
\hline
\noalign{\smallskip}
\endfoot

\noalign{\smallskip}
\hline
\noalign{\smallskip}
\endlastfoot

2M J10463795$-$7736035 &        F0V &       1 &    $\mu$,$\pi$ &        2 \\
2M J10523694$-$7440287 &      M4.75 &        3 &        NaK &        3 \\
2M J10533978$-$7712338 &      M2.75 &       4 &      $A_V$,ex &       4 \\
2M J10550964$-$7730540 &    M5,M4.5 &    5,3 &        NaK &    5,3 \\
T3B &       M3.5 &        3 &        NaK &        3 \\
2M J10555973$-$7724399 & M1,M0.5,M0.5,K8-M0.5 & 6,7,8,3 & e,ex,rv,$A_V$ & (9,10,6,7,8,11,12,13,3),(14,15,16),17,3 \\
2M J10561638$-$7630530 &       M5.6 &        3 &     NaK,e &        3 \\
2M J10563044$-$7711393 & K2,M0:,M0.5 & 18,8,19 & e,ex,Li,rv & (9,10,18,6,7,8,12,19),(14,15,16),20,17 \\
2M J10574219$-$7659356 &      M3.25 &        19 & e,ex,rv,NaK & (10,12,19),15,21,19 \\
2M J10580597$-$7711501 &      M5.25 &        3 &        NaK &        3 \\
2M J10581677$-$7717170 &         K0 &        7 &      e,ex & (7,22,12),(14,15,16,23,24) \\
2M J10590108$-$7722407 & K0:,M0:,K8 & 6,8,19 & e,ex,Li,rv,$A_V$ & (9,10,18,6,7,8,12,19),(14,15,16),20,17,19 \\
2M J10590699$-$7701404 & G8,K0,K2,K2 & 9,18,7,8 &   e,ex,rv & (9,10,18,7,8,12,13),(14,15,16,23),21 \\
2M J11004022$-$7619280 &      M3.75 &        19 &     e,NaK & (9,10,12,19),19 \\
2M J11011370$-$7722387 &      M5.25 &        3 &        NaK &        3 \\
2M J11011875$-$7627025 &      M2.25 &        19 &         Li &        25 \\
2M J11011926$-$7732383A &      M7.25 &       26 &     NaK,$A_V$ &       26 \\
2M J11011926$-$7732383B &      M8.25 &       26 &        NaK &       26 \\
2M J11013205$-$7718249 &    M8.5,M8 &    5,3 &        NaK &    5,3 \\
2M J11020610$-$7718079 &         M8 &       4 &  e,NaK,ex &       4 \\
2M J11021927$-$7536576 &       M4.5 &        3 &        NaK &        3 \\
2M J11022491$-$7733357 & K2,K2,K5,M0,K6 & 9,18,7,8,19 & e,ex,Li,rv & (9,10,18,7,8,11,12,19),(14,15,16,23,24),20,21 \\
2M J11022610$-$7502407 &      M4.75 &        3 &        NaK &        3 \\
2M J11023265$-$7729129 &         M3 &        19 & ($A_V$,Li),ex &    19,24 \\
2M J11024183$-$7724245 &         M5 &        3 &     NaK,$A_V$ &        3 \\
2M J11025374$-$7722561 &       M8.5 &       4 &     NaK,ex &       4 \\
2M J11025504$-$7721508 &  M0.5,M4.5 &   16,19 &  e,ex,NaK & (9,11,16,12,27,19),(16,24),19 \\
2M J11034186$-$7726520 &       M5.5 &        19 &        NaK &        19 \\
2M J11034764$-$7719563 &         M5 &        19 &     NaK,$A_V$ &        19 \\
2M J11035682$-$7721329 &    M4,M3.5 &    25,19 &     Li,NaK &    25,19 \\
2M J11040425$-$7639328 & M4.25,M4.5 &    19,5 & (e,NaK,$A_V$),ex & (19,5),24 \\
2M J11040909$-$7627193 &  K2,K7:,K5 & 6,8,19 & e,ex,Li,rv & (9,18,6,7,8,12,13,19),(15,16,24),20,(21,17) \\
2M J11041060$-$7612490 &         M6 &        19 &        NaK &        19 \\
2M J11042275$-$7718080 &      K6-M0 &        19 &      e,ex & (12,19),(15,27,24) \\
2M J11044258$-$7741571 &    M4,M4.5 &    19,5 &     ex,NaK & (27,24),(19,5) \\
2M J11045100$-$7625240 &   K0-K3,K8 &    25,19 &         Li &        25 \\
2M J11045285$-$7625514 & F8-G5,M1.75 &    25,19 &         Li &        25 \\
2M J11045701$-$7715569 &         M3 &        19 &      ex,$A_V$ & (27,24),19 \\
2M J11050752$-$7812063 &      M4.75 &        3 &        NaK &        3 \\
2M J11051467$-$7711290 &      M3.25 &        19 &     $A_V$,NaK &        19 \\
2M J11052272$-$7709290 &      M4.75 &        3 &     NaK,$A_V$ &        3 \\
2M J11052472$-$7626209 &      M2.75 &        3 &        NaK &        3 \\
2M J11054300$-$7726517 &      M5.25 &        19 &        NaK &        19 \\
2M J11055261$-$7618255 &       M1.5 &        19 &         Li &        19 \\
2M J11055780$-$7607488 &   B6IV/V,B7V &  1,28 &    $\mu$,$\pi$ &        2 \\
2M J11060010$-$7507252 &       M4.5 &        3 &        NaK &        3 \\
2M J11061540$-$7721567 &      G3-G7 &       29 &   e,ex,rv & 30,(31,15,32),33 \\
2M J11061545$-$7737501 &   M2,M2.75 &   34,19 &         $A_V$ &        19 \\
2M J11062554$-$7633418 &   M6,M5.25 &    5,3 & (NaK,e,$A_V$),ex & (5,3),24 \\
CHXR73B &       M9.5 &        35 &     H$_2$O,$A_V$ &        35 \\
Cha J11062854$-$7618039 &      M8-L0 &       4 &  e,NaK,ex &       4 \\
2M J11062877$-$7737331 & M4.5:,M3.25 &  36,19 &     $A_V$,NaK &        19 \\
2M J11062942$-$7724586 &         M6 &        24 &  $A_V$,H$_2$O,ex &        24 \\
2M J11063276$-$7625210 &         M6 &        19 & (e,NaK),ex &    19,24 \\
Cha$-$MMS1 &         ~$\cdots$~&         ~$\cdots$~&         ex &  37,32 \\
2M J11063799$-$7743090 &   M7:,M6.5 &  38,19 &  Li,rv,NaK & 17,17,19 \\
2M J11063945$-$7736052 &   M3,M5.25 &   39,19 &  ex,NaK,$A_V$ & (27,24),19,19 \\
2M J11064180$-$7635489 &       M4.5 &        19 &  e,NaK,ex & (12,19),19,24 \\
2M J11064346$-$7726343 &         M3 &        19 &      Li,$A_V$ &        19 \\
2M J11064510$-$7727023 &         K6 &        19 &   ex,Li,$A_V$ & (14,31,27,24),(40,19),19 \\
2M J11064658$-$7722325 &         ~$\cdots$~&         ~$\cdots$~&         ex & 31,15,23,27,32,24 \\
2M J11065733$-$7742106 & M4.5,M4.25 &  36,19 &  Li,rv,NaK & (41,17),17,19 \\
2M J11065803$-$7722488 &         ~$\cdots$~&         ~$\cdots$~&         ex & 27,32,24 \\
2M J11065906$-$7718535 & M1.5,M4.25 &   16,3 &  e,ex,NaK & (9,10,16,12),(31,15,27,24),3 \\
2M J11065939$-$7530559 &      M5.25 &        3 & (e,NaK),ex &    3,24 \\
2M J11070324$-$7610565 &         M6 &        3 &        NaK &        3 \\
ESOH$\alpha$281 &       M4.5 &       42 &      Li,ex &   42,24 \\
2M J11070369$-$7724307 &       M7.5 &        24 &  $A_V$,H$_2$O,ex &        24 \\
Cha J11070768$-$7626326 &         L0 &   24 &      e,ex &   24 \\
2M J11070919$-$7723049 &         ~$\cdots$~&         ~$\cdots$~&         ex & 43,31,15,27,32,24 \\
2M J11070925$-$7718471 & M2.5,M2-M4 &   34,3 &         ex &    27,24 \\
2M J11071148$-$7746394 &         M3 &        19 &     $A_V$,NaK &        19 \\
2M J11071181$-$7625501 &      M5.25 &        19 & (NaK,$A_V$),ex &    19,24 \\
2M J11071206$-$7632232 &       M0.5 &        19 & e,ex,Li,$A_V$ & (10,12,19),(15,16,24),19,19 \\
2M J11071330$-$7743498 &       M3.5 &        19 & ($A_V$,NaK),ex &    19,24 \\
2M J11071622$-$7723068 &         M1 &       34 &         ex &    27,24 \\
2M J11071668$-$7735532 & M7.5-M8,M7.5,M6-M7,M7.75 & 36,38,39,19 & e,ex,NaK,Li,rv & 36,(27,44),38,(41,17),17 \\
2M J11071860$-$7732516 & M5,M6,M5.5 & 39,38,19 & e,ex,NaK,$A_V$ & (39,38,19),(27,24),(38,19),19 \\
2M J11071915$-$7603048 & M2,M3,M2.5 & 7,8,19 &      Li,$A_V$ &        19 \\
2M J11072022$-$7738111 &      M4.25 &        3 &     NaK,$A_V$ &        3 \\
2M J11072040$-$7729403 &       M4.5 &        3 &        NaK &        3 \\
2M J11072074$-$7738073 & G2,G2,G2:,G2,G2 & 9,18,7,8,33 &   e,rv,ex & (9,10,18,7,8,12,33),(33,21),(14,45,15,16,23,27,24) \\
2M J11072142$-$7722117 &         M2 &       34 &      ex,e & (27,24),34 \\
2M J11072443$-$7743489 &      M5.75 &        3 &     NaK,$A_V$ &        3 \\
Cha J11072647$-$7742408 &       $\geq$M9 &   3 &        H$_2$O &   3 \\
2M J11072825$-$7652118 & K0:,M1,M1.5,M1.5+M3,M3 & 6,7,8,42,3 & e,ex,rv,Li & (9,10,6,7,8,12,42,3),(15,16,24),21,42 \\
2M J11073519$-$7734493 &   M5,M4.25 &  36,19 & Li,rv,NaK,$A_V$,ex & (41,17),17,19,19,24 \\
2M J11073686$-$7733335 &       M3.5 &        19 &     NaK,$A_V$ &        19 \\
2M J11073775$-$7735308 &   M8,M7.75 &  38,19 & NaK,Li,rv,e & (38,19),41,17,19 \\
2M J11073832$-$7747168 &    M5,M4.5 &    5,3 &     NaK,$A_V$ &  (5,3) \\
2M J11073840$-$7552519 &      M4.75 &        3 &        NaK &        3 \\
2M J11074245$-$7733593 & M6,M6.5,M5,M5.25 & 36,38,39,19 & e,ex,NaK,Li,rv,$A_V$ & (36,19),(27,24),(38,19),41,17,19 \\
2M J11074366$-$7739411 & M0.5,M0.5,M1,M0 & 7,16,36,3 &   e,ex,$A_V$ & (9,10,7,12,36,3),(14,15,27,24),3 \\
2M J11074610$-$7740089 & M6.5,M5.75 &  38,19 &  NaK,Li,rv & (38,19),(41,17),17 \\
2M J11074656$-$7615174 & M6.5,M5.75 &   34,19 &     NaK,ex &    19,24 \\
2M J11075225$-$7736569 & M6,M7,M5.5 & 36,38,19 & NaK,Li,rv,$A_V$ & (38,19),(41,17),17,19 \\
2M J11075588$-$7727257 &      K7,K5 &   29,33 &      Li,rv &        33 \\
2M J11075730$-$7717262 & M0.75-M1.75 &        19 &   ex,e,$A_V$ & (14,43,31,15,16,27,24),19,19 \\
2M J11075792$-$7738449 &   K7,K4-K6 &  36,3 &   e,ex,$A_V$ & (9,10,12,36,3),(14,15,27,24),3 \\
2M J11075809$-$7742413 &  M2.5,M2.5 &  36,3 & e,Li,rv,$A_V$,ex & (10,12,36,41,3),(41,17),17,3,24 \\
2M J11075993$-$7715317 & M6.5,M5.75 &    5,3 &     NaK,$A_V$ & (5,3),3 \\
2M J11080002$-$7717304 &         K8 &        19 &      ex,$A_V$ & (14,43,31,15,16,27,24),19 \\
2M J11080148$-$7742288 & K2,K5-K7+K7,M0.5,K8 & 8,42,36,19 & e, & (9,10,18,6,7,8,11,12,42,36,13,19), \\
 & &  & ex,Li,rv,$A_V$ & (14,45,15,16,27,24),(20,42),17,19 \\
2M J11080234$-$7640343 &         M6 &        3 &        NaK &        3 \\
2M J11080297$-$7738425 & M2,M1.25,K: & 34,19,5 &   ex,e,$A_V$ & (27,24),(46,19,5),19 \\
2M J11080329$-$7739174 & B9.5,B9.5,B9-A0 & 9,7,11 &      e,ex & (9,7,11,12,30),(15,16,23,27,24) \\
2M J11080609$-$7739406 &       $\geq$M9 &   3 &        H$_2$O &   3 \\
T33B &   K3,K7,K6 & 29,36,19 &   e,Li,$A_V$ & (29,19),19,19 \\
2M J11081509$-$7733531 & G3-G7,G3-K3 &   29,19 & e,ex,Li,$A_V$ & (29,12,30,13,19),(14,45,15,16,23,27,24),(20,19),19 \\
2M J11081648$-$7744371 & M3.5,M3.75 &   16,19 &        NaK &        19 \\
2M J11081703$-$7744118 &    M5,M5.5 &  38,19 &        NaK &        19 \\
2M J11081850$-$7730408 &  M5.5,M6.5 &   39,19 &  ex,e,NaK & (27,44),39,19 \\
2M J11081896$-$7739170 & M6.5,M6,M5.5 & 36,38,19 &  NaK,Li,rv & (38,19),(41,17),17 \\
Cha J11081938$-$7731522 &      M4.75 &        3 &     NaK,ex &    3,24 \\
2M J11082238$-$7730277 &    M5.5,M5 &   39,19 & ex,e,NaK,$A_V$ & (27,24),(39,19),19,19 \\
2M J11082404$-$7739299 & M7.5,M6.25 &  38,19 &        NaK &  38,19 \\
2M J11082410$-$7741473 & M6,M6,M5.5 & 36,38,19 & NaK,Li,rv,$A_V$ & (38,19),(41,17),17,19 \\
2M J11082570$-$7716396 &         M8 &        3 &     NaK,ex &  3,44 \\
2M J11082650$-$7715550 &      M5.75 &        19 &  NaK,$A_V$,ex & 19,19,24 \\
2M J11082927$-$7739198 &   M8,M7.25 &  38,19 &     NaK,ex & 38,19,44 \\
Cha J11083040$-$7731387 &       $\geq$M9 &   3 &        H$_2$O &   3 \\
2M J11083896$-$7743513 &        $<$M0 &        19 &         ex & 45,15,16,23,27,24 \\
2M J11083905$-$7716042 &      M0,K8 &   16,19 &   e,ex,$A_V$ & (9,10,16,12,19),(31,15,16,24),19 \\
2M J11083952$-$7734166 & M6,M7,M5.75 & 36,38,19 & Li,NaK,rv,e,ex & (41,17),(38,19),17,19,24 \\
2M J11084069$-$7636078 & M0,M0,M2.5 & 29,25,3 &      Li,$A_V$ &    25,3 \\
2M J11084296$-$7743500 &      M3-M5 &        24 &   $A_V$,e,ex &        24 \\
2M J11084952$-$7638443 &      M8.75 &        3 &  e,NaK,ex &  3,44 \\
2M J11085090$-$7625135 &      M5.25 &        19 &     NaK,ex &    19,24 \\
2M J11085176$-$7632502 &      M7.25 &        3 &     NaK,$A_V$ &        3 \\
2M J11085242$-$7519027 &      M2,M3 &    47,48 &   Li,$\mu$,ex & 48,24,24 \\
2M J11085326$-$7519374 &      K6,K7 &    47,48 &      Li,$\mu$ &    48,24 \\
2M J11085367$-$7521359 &       M1.5 &       4 &      e,ex &       4 \\
2M J11085421$-$7732115 & M5.5,M5.25 &  36,19 &     Li,NaK &   41,19 \\
2M J11085464$-$7702129 & M0.5:,M0.5 &    8,19 &   e,ex,$A_V$ & (9,10,8,11,12,19),(15,27,24),19 \\
2M J11085497$-$7632410 &    M5.5,M6 &    19,5 & (e,$A_V$,NaK),ex & (19,5),24 \\
2M J11085596$-$7727132 &      M5.25 &        3 &     NaK,$A_V$ &        3 \\
2M J11090512$-$7709580 &      M4.75 &        19 &        NaK &        19 \\
2M J11091172$-$7729124 & M0,M0.5+M2,M2 & 16,42,19 &         Li &       42 \\
T39B &         M3 &        3 &         Li &        3 \\
2M J11091297$-$7729115 &         M3 &       4 &        NaK &       4 \\
Cha J11091363$-$7734446 &       M9.5 &    49,24 & (H$_2$O,ex),NaK &    49,24 \\
2M J11091380$-$7628396 &      M4.75 &        19 &        NaK &        19 \\
2M J11091769$-$7627578 &   K7,K7,K7 & 47,48,3 &   Li,rv,$A_V$ & (25,47,48),48,3 \\
2M J11091812$-$7630292 & M0.75-M1.75 &        19 &   e,ex,$A_V$ & (12,19),(27,24),19 \\
2M J11092266$-$7634320 & M0.75-M1.75 &        19 &   ex,e,$A_V$ & (45,15,27,24),19,19 \\
2M J11092379$-$7623207 &      K6,K6 &    8,3 & e,ex,Li,rv,$A_V$ & (9,10,18,6,7,8,11,12,13,3),(45,15,16,24),20,17,3 \\
2M J11092855$-$7633281 &      M6.5? &       34 &      e,ex &   34,24 \\
2M J11092913$-$7659180 &      M5.25 &        3 &     NaK,$A_V$ &        3 \\
2M J11093543$-$7731390 &      M8.25 &        3 &        NaK &        3 \\
2M J11093777$-$7710410 &      M2,K7 &   34,19 &      Li,$A_V$ &        19 \\
2M J11094006$-$7628391 &   M1,M1.25 &   29,19 &      Li,$A_V$ & (25,19),19 \\
2M J11094192$-$7634584 &        $<$M0 &        3 &         ex &    27,24 \\
2M J11094260$-$7725578 &      M5,M6 &    19,5 & ex,e,NaK,$A_V$ & (31,15,27,24),(19,5),(19,5),(19,5) \\
2M J11094525$-$7740332 & M6.5,M5.75 &    5,3 &     NaK,$A_V$ &  (5,3) \\
2M J11094621$-$7634463 &      M3.25 &        19 & e,ex,$A_V$,NaK & (9,10,12,19),(27,24),19,19 \\
2M J11094742$-$7726290 &   M3.25,M1 &    19,5 & ex,e,$A_V$,NaK & (27,24),(19,5),(19,5),19 \\
2M J11094866$-$7714383 &         M1 &       39 &         ex &    27,24 \\
2M J11094918$-$7731197 & M6,M5.5,M5.5 & 46,19,3 &     e,NaK & 46,(46,3) \\
2M J11095003$-$7636476 &         B9 &        7 &   e,$A_V$,ex & 30,7,(14,45,15,16,27) \\
2M J11095215$-$7639128 &      M6-M7 &        19 & e,$A_V$,NaK,ex &  19,44 \\
2M J11095262$-$7740348 &         M3 &        19 &      e,ex &    19,24 \\
2M J11095336$-$7728365 & M4.5,M5.5,M5.75 & 39,46,19 & ex,e,NaK,$A_V$ & (27,24),(46,19),(46,19),19 \\
2M J11095340$-$7634255 &      K4-K6 &        19 &   e,ex,$A_V$ & (12,19),(15,16,23,27,24),19 \\
2M J11095407$-$7629253 &   M0,M2,M2 & 39,46,19 &   e,ex,$A_V$ & (10,12,39,46,19),(27,24),19 \\
2M J11095437$-$7631113 & M2,M3,M1.75 & 39,46,19 &      ex,e & (27,24),(46,19) \\
2M J11095493$-$7635101 &      M5.75 &       4 & e,$A_V$,NaK,ex &       4 \\
2M J11095505$-$7632409 &        $<$M0 &        3 &      ex,e & (27,24),3 \\
2M J11095873$-$7737088 & K0:,K7-M0,K7-M0,M1.25 & 6,7,8,3 &      e,ex & (9,10,6,7,8,12,13,3),(14,45,15,16,27,24) \\
2M J11100010$-$7634578 & $<$K6,K5:,$<$K6 & 7,8,19 &   e,ex,$A_V$ & (9,10,18,7,8,12,13,19),(15,16,23,27,24),19 \\
2M J11100192$-$7725451 &      M5.25 &        3 &     NaK,$A_V$ &        3 \\
2M J11100336$-$7633111 &      M3-M5 &       4 &   e,ex,$A_V$ &       4 \\
2M J11100369$-$7633291 &         K8 &        19 &   e,ex,$A_V$ & (12,19),(27,24),19 \\
2M J11100469$-$7635452 &      M0,M1 &    8,3 &      e,ex & (8,3),(27,24) \\
2M J11100658$-$7642486 &      M9.25 &        3 &        NaK &        3 \\
2M J11100704$-$7629376 & K2:,K7,M0,M0 & 6,7,8,3 & e,ex,rv,$A_V$ & (9,10,18,6,7,8,11,12,3),(45,15,16,27,24),(21,17),3 \\
2M J11100785$-$7727480 &    M5,M5.5 &   39,19 &  ex,NaK,$A_V$ & (27,24),19,19 \\
2M J11100934$-$7632178 & $\geq$M9.5,M9.5 &    50,3 & H$_2$O,ex,NaK & 50,51,3 \\
2M J11101141$-$7635292 & M1.5,M0,K5.5 & 34,46,19 &   ex,e,$A_V$ & (27,24),19,19 \\
2M J11101153$-$7733521 &       M4.5 &        3 &     NaK,$A_V$ &        3 \\
2M J11102226$-$7625138 &         M8 &        19 &     e,NaK &        19 \\
2M J11102852$-$7716596 &       M5.5 &        19 &        NaK &        19 \\
2M J11103481$-$7722053 &         M4 &        19 &     NaK,$A_V$ &        19 \\
2M J11103644$-$7722131 &   M5,M4.75 &   39,19 &     NaK,$A_V$ &        19 \\
2M J11103801$-$7732399 &      K3,K3 &    11,48 &   ex,Li,rv & (14,45,43,15,27,24),(20,48),48 \\
2M J11104006$-$7630547 &      M7.25 &        3 &     NaK,$A_V$ &        3 \\
2M J11104141$-$7720480 &      M2,M6 &   39,19 &  ex,e,NaK & (27,24),(39,19),19 \\
2M J11104959$-$7717517 &     K7:,M2 &    8,19 &   e,ex,$A_V$ & (9,10,8,12,19),(15,27,24),19 \\
2M J11105076$-$7718031 & M4.5,M4.25 &    5,3 &     NaK,$A_V$ &  (5,3) \\
2M J11105333$-$7634319 & M1,M1:,M3.75 & 7,8,3 &      e,ex & (9,10,18,6,7,8,11,12,27,3),(15,24) \\
2M J11105359$-$7725004 & M5,M5,M4.5 & 39,34,19 & ex,e,NaK,$A_V$ & (27,24),(39,19),19,19 \\
2M J11105597$-$7645325 &      M5.75 &        19 &     ex,NaK & (15,27,24),19 \\
2M J11111083$-$7641574 &   K7:,M2.5 &    5,3 &   e,$A_V$,ex & (5,3),3,24 \\
2M J11112249$-$7745427 &      M8.25 &        3 &     NaK,ex &  3,44 \\
2M J11112260$-$7705538 &       M4.5 &        19 &     NaK,$A_V$ &        19 \\
2M J11113474$-$7636211 & K5-M0,M2.5 &    25,19 &         Li &        25 \\
2M J11113965$-$7620152 & M0,M2,M1.5-M2.5 & 7,8,19 &      e,ex & (9,10,6,7,8,11,12,13,19),(15,16,24) \\
2M J11114533$-$7636505 &         M8 &        3 &     NaK,ex &  3,44 \\
2M J11114632$-$7620092 & K1,K3,K3,K6 & 33,47,48,19 & rv,e,Li,ex & (33,48),(33,47,48,19),(33,12,47,48),24 \\
2M J11115400$-$7619311 & M1,K7-M0,M2.5 & 12,25,3 &         Li &        25 \\
2M J11120288$-$7722483 &         M6 &        3 &     $A_V$,NaK &        3 \\
2M J11120327$-$7637034 &       M5.5 &        19 &        NaK &        19 \\
2M J11120351$-$7726009 & M3.5,M4.75,M6 & 39,19,5 & ex,e,NaK,$A_V$ & (27,24),19,(19,5),(19,5) \\
2M J11120984$-$7634366 &         M5 &        19 &  e,NaK,ex &    19,24 \\
2M J11122250$-$7714512 &      M9.25 &        3 &     NaK,ex &    3,24 \\
2M J11122441$-$7637064 &      K3,K4 &   29,48 & e,Li,ex,rv & 29,48,(45,15,16,24),(21,48) \\
2M J11122772$-$7644223 & G8,G9,G8:,K0 & 9,18,7,8 &   e,ex,rv & (9,10,18,7,8,12,30),(45,15,16,24),(21,17) \\
2M J11123092$-$7644241 &    M0.5,M1 &   16,19 &   e,$A_V$,ex & (9,10,16,12,19),19,24 \\
2M J11123099$-$7653342 &         M7 &        3 &        NaK &        3 \\
2M J11124210$-$7658400 &   K7,K5,M1 & 29,33,19 & rv,Li,e,$A_V$ & 33,33,19,19 \\
2M J11124268$-$7722230 & K0,G5-K0,G5 & 7,29,33 &   rv,Li,ex & 33,33,(14,15,16,24) \\
2M J11124299$-$7637049 &      K5,K4 &    33,47 &      rv,Li & 33,(33,47) \\
2M J11124861$-$7647066 &         M4 &        19 & (Li,NaK),ex &    19,24 \\
2M J11132012$-$7701044 & K0-M0,M2.75 &    25,19 &         Li &    25,19 \\
2M J11132446$-$7629227 &       M3.5 &        19 &     NaK,ex &    19,24 \\
2M J11132737$-$7634165 &   M2.75 &    19 &         Li &    25,19 \\
2M J11132970$-$7629012 &      M4.25 &        19 &     Li,NaK &        19 \\
2M J11133356$-$7635374 &       M4.5 &        19 &        NaK &        19 \\
2M J11141565$-$7627364 &      M3.75 &        19 &        NaK &        19 \\
2M J11142454$-$7733062 &         M4 &        19 & e,NaK,$A_V$,ex & (12,19),19,19,24 \\
2M J11142611$-$7733042 &      M5.75 &        19 & (NaK,$A_V$),ex &    19,24 \\
2M J11142906$-$7625399 & M5.5,M4.75 &    5,3 &        NaK &    5,3 \\
2M J11145031$-$7733390 &      M2.75 &        19 &         Li &        19 \\
2M J11152180$-$7724042 & M5.5,M4.75 &    5,3 &        NaK &    5,3 \\
2M J11155827$-$7729046 &       M4.5 &        3 &        NaK &        3 \\
2M J11160287$-$7624533 &   K:,K7-M0 &    5,3 &         e &    5,3 \\
2M J11173700$-$7704381 &  M0.5,M0.5 &    8,19 &      e,ex & (9,10,6,7,8,12,19),(15,16,23) \\
2M J11173792$-$7646193 &      M5.75 &        3 &        NaK &        3 \\
2M J11175211$-$7629392 &       M4.5 &        3 &        NaK &        3 \\
2M J11181957$-$7622013 &      M2.25 &        19 &         Li &        19 \\
2M J11182024$-$7621576 &      K8 &    19 &         Li &    25,19 \\
2M J11183379$-$7643041 &         M5 &        3 &        NaK &        3 \\
2M J11194214$-$7623326 &         M5 &        3 &        NaK &        3 \\
2M J11195652$-$7504529 &      M7.25 &        3 &        NaK &        3 \\
2M J11241186$-$7630425 &         M5 &        3 & (NaK,$A_V$),ex &    3,24 \\
2M J11242980$-$7554237 &      M4.75 &        3 &        NaK &        3 \\
2M J11291261$-$7546263 &         K3 &    47,48 &      Li,$\mu$ &    48,24 \\
2M J11332327$-$7622092 &       M4.5 &        3 &        NaK &        3 \\

\end{longtable}
\vspace{-5mm}
}
\end{center}
\end{landscape}

\begin{figure}[tbp]
\centering
\vspace{-3mm}
\includegraphics[draft=False,width=0.45\textwidth]{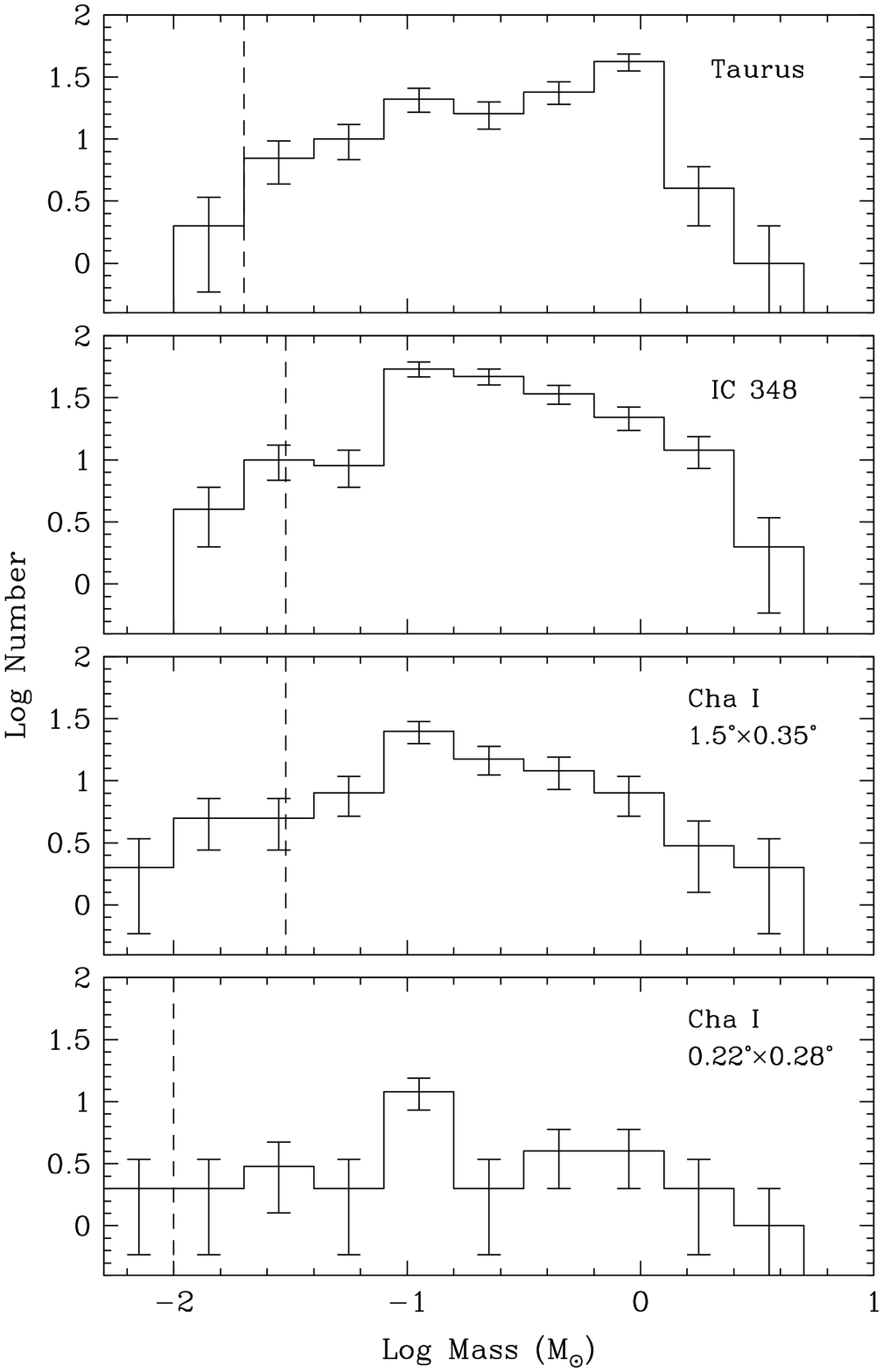}
\includegraphics[draft=False,width=0.45\textwidth]{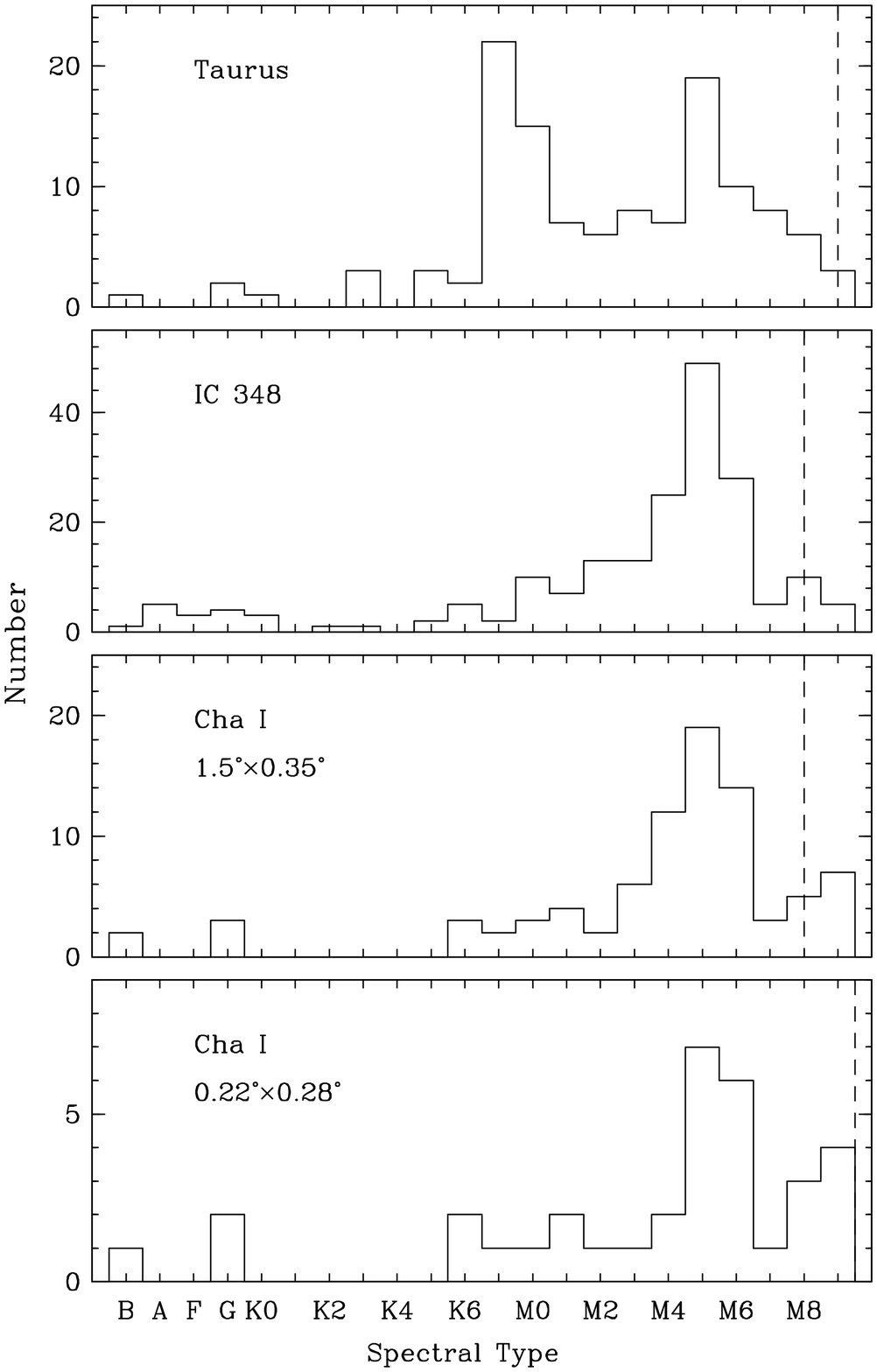}
\vspace{-3mm}
\caption{
{\it Left}: IMFs from \citet{luh07cha} for extinction-limited samples in
a field encompassing most of the Cha~I cloud ($1\fdg5\times0\fdg35$)
and in a small area in the southern cluster of Cha~I
($0\fdg22\times0\fdg28$) compared to IMFs in Taurus \citep{luh04tau}
and IC~348 \citep{luh03b}.
In the units of this diagram, the Salpeter slope is 1.35.
{\it Right}: Distributions of spectral types in the IMFs.
The completeness limits of these samples are indicated ({\it dashed lines}).
}
\label{fig:imf}

\includegraphics[draft=False,scale=0.4]{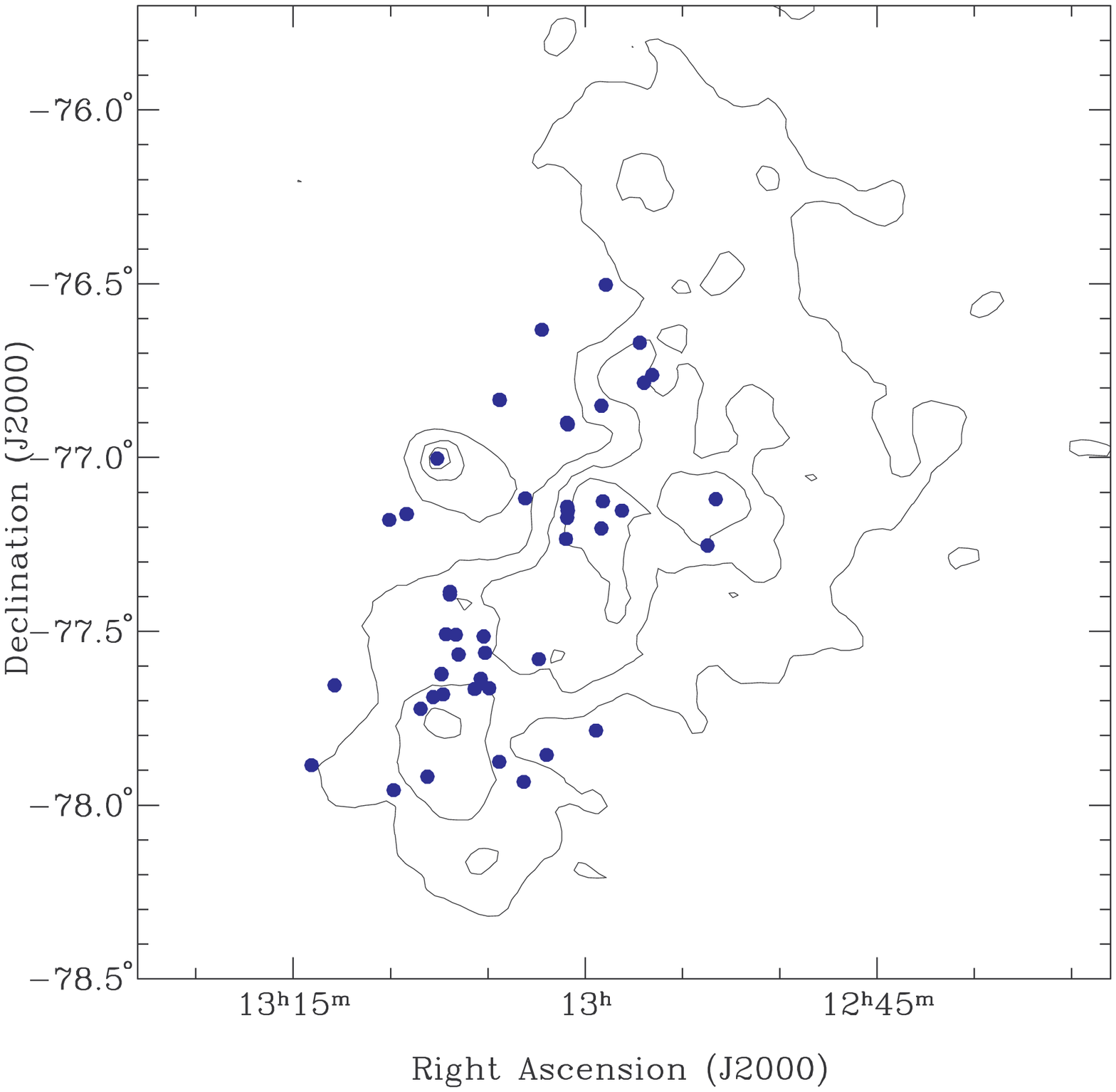}
\vspace{-2mm}
\caption{
Spatial distribution of known members of Cha~II
\citep[{\it points},][]{alc06,all07,spe07,spe08}.
The contours represent the 100~\micron\ map from {\it IRAS}.
}
\label{fig:mapII}
\end{figure}

\begin{table}
\begin{center}
\caption{References for Table 2}
\smallskip
\small
{\footnotesize
\begin{tabular}{cccc}
\tableline
\noalign{\smallskip}
Number & Reference & Number & Reference \\
\noalign{\smallskip}
\tableline
\noalign{\smallskip}
1 & \citet{hc75} & 27 & \citet{per00} \\
2 & \citet{per97} & 28 & \citet{vr84} \\
3 & \citet{luh07cha} & 29 & \citet{fk89} \\
4 & \citet{luh08cha2} & 30 & \citet{gur99} \\
5 & \citet{com04} & 31 & \citet{pru91} \\
6 & \citet{app79} & 32 & \citet{leh01} \\
7 & \citet{ryd80} & 33 & \citet{wal92} \\
8 & \citet{app83} & 34 & \citet{gm03} \\
9 & \citet{hm73} & 35 & \citet{luh06bin} \\
10 & \citet{sch77} & 36 & \citet{com99} \\
11 & \citet{whi87} & 37 & \citet{rei96} \\
12 & \citet{har93} & 38 & \citet{com00} \\
13 & \citet{nat00} & 39 & \citet{gp02} \\
14 & \citet{bau84} & 40 & \citet{law96} \\
15 & \citet{pru92b} & 41 & \citet{nc99} \\
16 & \citet{gs92} & 42 & \citet{bz97} \\
17 & \citet{jg01} & 43 & \citet{whi91b} \\
18 & \citet{app77} & 44 & \citet{luh05frac} \\
19 & \citet{luh04cha} & 45 & \citet{ass90} \\
20 & \citet{gre92} & 46 & \citet{saf03} \\
21 & \citet{dub96} & 47 & \citet{alc95} \\
22 & \citet{whi97} & 48 & \citet{cov97a} \\
23 & \citet{hen93} & 49 & \citet{luh05disk2} \\
24 & \citet{luh08} & 50 & \citet{luh04ots} \\
25 & \citet{hlf94} & 51 & \citet{luh05ots} \\
26 & \citet{luh04bin} &  &  \\
\noalign{\smallskip}
\tableline
\end{tabular}
}
\end{center}
\end{table}

\noindent of the census of known members and the molecular observations
of the cloud imply a star formation efficiency of $\sim10$\% for Cha~I,
which is an upper limit since some of the cloud has probably dissipated,
particularly in the northern half \citep{luh07cha}.
The initial mass functions (IMFs) measured for two fields in Cha~I
are shown in Figure~\ref{fig:imf} \citep{luh07cha}. The IMF in Cha~I reaches
a maximum at a mass of 0.1-0.15~$M_\odot$, which is similar to the turnover
mass observed in IC~348 and the Orion Nebula Cluster
\citep{hil97,hc00,mue02,mue03,luh03b}.
The substellar IMF is roughly flat in logarithmic units and shows no indication
of reaching a minimum down to a completeness limit of 0.01~$M_\odot$.

Spectral types and evidence of membership are less complete for Cha~II
than for Cha~I. \citet{spe07,spe08} and \citet{alc08} claimed that their
census of members of Cha~II is complete to 0.03~$M_\odot$ for $A_V<2$.
However, they only demonstrated that their photometry is complete
to that mass limit, and did not show that their spectroscopic survey
is complete to that limit. Indeed, 11 candidate members identified
by \citet{alc08} and \citet{spe08} lack the spectroscopy needed to assess
membership. Compilations of $\sim$50 likely members of Cha~II have
been published by \citet{alc00}, \citet{you05}, and \citet{spe07,spe08}.
Spectroscopic confirmation of individual
late-type members of Cha~II has been presented by \citet{bar04},
\citet{alc06}, and \citet{all07}. In Figure~\ref{fig:mapII}, the positions of
the known members of Cha~II are plotted with the 100~\micron\ {\it IRAS} map
of the cloud.  No young stars have been found in Cha~III.
Using data from the Hipparcos catalog,
\citet{sar03} identified 21 B- and A- type stars with distances
between 100 and 200~pc and suggested that these stars comprise an OB
association that is associated with the Chamaeleon clouds.

\section{Multiplicity}

Members of Chamaeleon have been targeted in several multiplicity surveys.
\citet{rz93} and \citet{bra96} used ground-based direct imaging at
$\sim$1~$\mu$m to search for wide binaries among nearly 200 stars across the
Chamaeleon cloud
complex, including members of the widely distributed X-ray population
\citep{alc95}. \citet{bz97} obtained resolved spectroscopy and photometry
for components of a sample of double stars found in those and other studies.
The components of the pairs were roughly coeval when placed on the
H-R diagram and compared to theoretical isochrones.
Using photometry from 2MASS, \citet{kra07} performed a deeper search
for wide companions in Cha~I that reached substellar masses.
They identified several candidate companions, some of which are discussed
by \citet{luh07cha}.
To detect binaries at smaller separations among members of Chamaeleon,
higher resolution imaging has been utilized, including
$K$-band speckle \citep{ghe97}, adaptive optics
\citep[AO,][]{bra01,ahm07,laf08}, and the {\it Hubble Space Telescope}
\citep{neu02}.
The latter study focused on the Cha~H$\alpha$ low-mass stars and brown dwarfs
\citep{com99,com00,nc99}. Because the primaries were intrinsically faint,
the images from \citet{neu02} were capable of detecting companions at very
low masses ($M_2>5$~$M_{\rm Jup}$ at separations greater than $0\farcs35$).
One promising candidate low-mass companion was identified, but it was later
classified as a background star through spectroscopy \citep{neu03}.
Like \citet{neu02}, \citet{ahm07} selected late-type members of Cha~I
for their AO survey, which consisted of 22 low-mass stars ($\leq$M6)
and 6 brown dwarfs ($>$M6). Their data resolved the low-mass stars
Cha~H$\alpha$~2, Hn~13, and CHXR~15 into doubles. Cha~H$\alpha$~2
was previously identified as a binary by \citet{neu02}.
Through proper motion measurements, \citet{sch08} demonstrated that neither of
the components of Cha~H$\alpha$~2 is a background star.
The most thorough multiplicity survey of Cha~I to date was performed by
\citet{laf08}, who obtained high-resolution images for 126
members with masses ranging from 0.1 to 3~$M_\odot$ (the targets from
\citet{ahm07} represented a subsample of this survey).

\begin{figure}
\centering
\includegraphics[draft=False,width=0.6\textwidth]{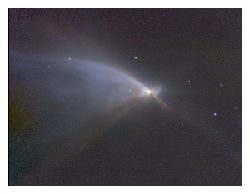}
\caption{
The IRN as seen in a $JHK$ color-composite image
\citep[$166\arcsec\times125\arcsec$,][]{zin99}. North is up and east is left.
}
\label{fig:irn}
\end{figure}

During spectroscopy of candidate young brown dwarfs in Cha~I,
\citet{luh04bin} serendipitously discovered a $1\farcs4$ pair of
objects, 2MASS J11011926$-$ 7732383~A and B, that are likely to
have substellar masses according to evolutionary models.
The projected angular separation of this system corresponds to 240~AU
at the distance of Cha~I, making it the first known
binary brown dwarf with a separation greater than 20~AU.
In another survey for free-floating brown dwarfs in Cha~I,
\citet{luh06bin} discovered a $1\farcs3$ companion to the low-mass star CHXR~73
using optical images obtained with the {\it Hubble Space Telescope}.
Based on its spectral type and luminosity, this companion appears to have
a mass of 10-15~$M_{\rm Jup}$.
Through $K$-band imaging of the area around Ced~110, \citet{per01}
resolved the protostar Ced~110~IRS6 as a $2\arcsec$ pair.
During a near- and mid-IR multiplicity survey of protostars in nearby
star-forming regions, \citet{hai04,hai06} found that both components
of Ced~110~IRS6 exhibit Class~I spectral energy distributions.
That survey also uncovered a $2\arcsec$ companion to ISO~97 that has a
steeply rising spectral energy distribution, possibly indicative of a
Class~0 source. In their {\it Spitzer} survey of Cha~I,
\citet{luh08} detected mid-IR excess emission from a faint source that is only
$14\arcsec$ from IRN (directly left of IRN in Figure~\ref{fig:irn}).
Through near-IR spectroscopy,
they classified it as a highly embedded young low-mass star ($A_J=5.6$, M4).
\citet{luh08} also detected a very faint, red source $7\arcsec$
from the low-mass member 2MASS J11062942$-$7724586 (Figure~\ref{fig:image})
that is either a Class~I brown dwarf or a galaxy.
A compilation of known visual doubles with projected separations less
than 5$\arcsec$ in Cha~I is provided in Table~4. Pairs with larger
separations are resolved by 2MASS and thus can be found in Tables~1 and 2
if both components are confirmed members. \citet{ghe97} reported that
CHXR~47 is a $1\farcs2$ binary based on seeing-limited images, but the
AO images from \citet{laf08} show that a pair of this kind is not present.

To detect tight binaries that are unresolved by direct imaging,
high-resolution spectroscopy has been applied to members of Chamaeleon.
For instance, surveys of a few dozen solar-mass young stars in and around
the Chamaeleon clouds by \citet{melo03} and \citet{gue07} identified
CHX18N, BF~Cha, CS~Cha, and RX~J1220.6$-$7539 as spectroscopic binaries.
\citet{jg01} measured radial velocities for a dozen low-mass stars and
brown dwarfs in Cha~I at multiple epochs, finding possible
variations due to low-mass unseen companions. Additional evidence of
radial velocity variations has been provided by newer observations
\citep{joe06a}, which comprise the first radial velocity monitoring survey
of young low-mass stars and brown dwarfs.
These observations have uncovered a 16-20~$M_{\rm Jup}$ companion in
a 1590~day orbit around the low-mass star Cha~H$\alpha$~8 \citep{joe07}.

\setlength{\tabcolsep}{1.4\deftabcolsep}
\begin{table}
\begin{center}
{Table 4. \hspace{4mm} Visual Binaries in Chamaeleon I ($<5\arcsec$) }
\smallskip
\small
{\footnotesize
\begin{tabular}{ccc}
\tableline
\noalign{\smallskip}
Name & Separation (arcsec) & Reference \\
\noalign{\smallskip}
\tableline
\noalign{\smallskip}
 2M~J11011926$-$7732383 & 1.44 & 1 \\
 2M~J11103481$-$7722053 & 0.06 & 2 \\
 B53 & 0.30 & 2 \\
 Ced~110~IRS6 & 1.95 & 3,4 \\
 Cha H$\alpha$ 2 & 0.16 & 5,6,7 \\
 CHXR9C & 0.96 & 8 \\
 CHXR15 & 0.30 & 6 \\
 CHXR26 & 1.40 & 9,2 \\
 CHXR28 & 1.87 & 8 \\
 CHXR30A & 0.46 & 2 \\
 CHXR37 & 0.08 & 2 \\
 CHXR40 & 0.15 & 2 \\
 CHXR47 & 0.18 & 2 \\
 CHXR49NE & 0.27 & 2 \\
 CHXR59 & 0.15 & 2 \\
 CHXR62 & 0.12 & 2 \\
 CHXR71 & 0.57 & 2 \\
 CHXR73 & 1.30 & 10 \\
 CHXR79 & 0.89 & 8,2 \\
 Hn4 & 0.21 & 2 \\
 Hn13 & 0.13 & 6 \\
 ISO97 & 2.05 & 11 \\
 ISO126 & 0.29 & 2 \\
 T3 & 2.22 & 12,2 \\
 T5 & 0.16 & 2 \\
 T14a & 2.35 & 13,4 \\
 T21 & 0.14 & 2 \\
 T26 & 4.6,4.9 & 12,14 \\
 T27 & 0.79 & 12,2 \\
 T31 & 0.72,0.66 & 8,14 \\
 T33 & 2.43 & 12,14,4,2  \\
 T39 & 1.1 & 12 \\
 T41 & 0.79 & 14,2 \\
 T43 & 0.80 & 2 \\
 T45 & 0.75 & 14,2 \\
 T46 & 0.12 & 2 \\
 T51AB & 1.98 & 12,14,2 \\
 T51BC & 0.1 & 15 \\
 T54 & 0.25 & 14,2 \\
\noalign{\smallskip}
\tableline
\noalign{\smallskip}

\end{tabular}
}
\end{center}
{\small
(1) \citet{luh04bin};
(2) \citet{laf08};
(3) \citet{per01};
(4) \citet{hai04};
(5) \citet{neu02};
(6) \citet{ahm07};
(7) \citet{sch08};
(8) \citet{bra96};
(9) \citet{luh04cha};
(10) \citet{luh06bin};
(11) \citet{hai06};
(12) \citet{rz93};
(13) \citet{sch77};
(14) \citet{ghe97};
(15) \citet{bra01}.
}
\end{table}
\setlength{\tabcolsep}{\deftabcolsep}

\section{Kinematics and Rotation}

A modest amount of work has been done on the kinematics and rotation
of members of Chamaeleon.
Radial velocities have been measured through surveys for
spectroscopic binaries \citep{rei02,melo03,gue07} and through followup
spectroscopy of X-ray sources in Cha~I \citep{wal92}, X-ray sources
scattered across the cloud complex \citep{cov97a}, and candidate low-mass
members of Cha~I \citep{nc99}. Dedicated kinematic studies have been performed
for both solar-mass stars \citep{dub96} and low-mass stars and brown dwarfs
\citep{jg01,joe06b}.
As for stellar rotation, measurements of v~sin~i have been reported for a
few dozen solar-mass stars \citep{fra88} and low-mass stars and brown dwarfs
\citep{moh05}, mostly in Cha~I. In some of the first measurements of periods
for young low-mass objects through photometric monitoring, \citet{joe03}
observed rotational periods of a few days for low-mass stars and brown dwarfs,
which agreed with the v~sin~i data for those objects \citep{jg01}.

\section{Herbig-Haro Objects and Outflows}

\begin{figure}[p]
\includegraphics[scale=5]{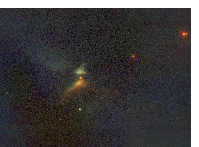}
\vspace{-2mm}
\caption{
The area surrounding Ced~110~IRS4 as seen in a $JHK$ color-composite image
\citep[$166\arcsec\times125\arcsec$,][]{zin99}. North is up and east is left.
}
\label{fig:irs4}

\vspace{-10mm}

\includegraphics[scale=0.4]{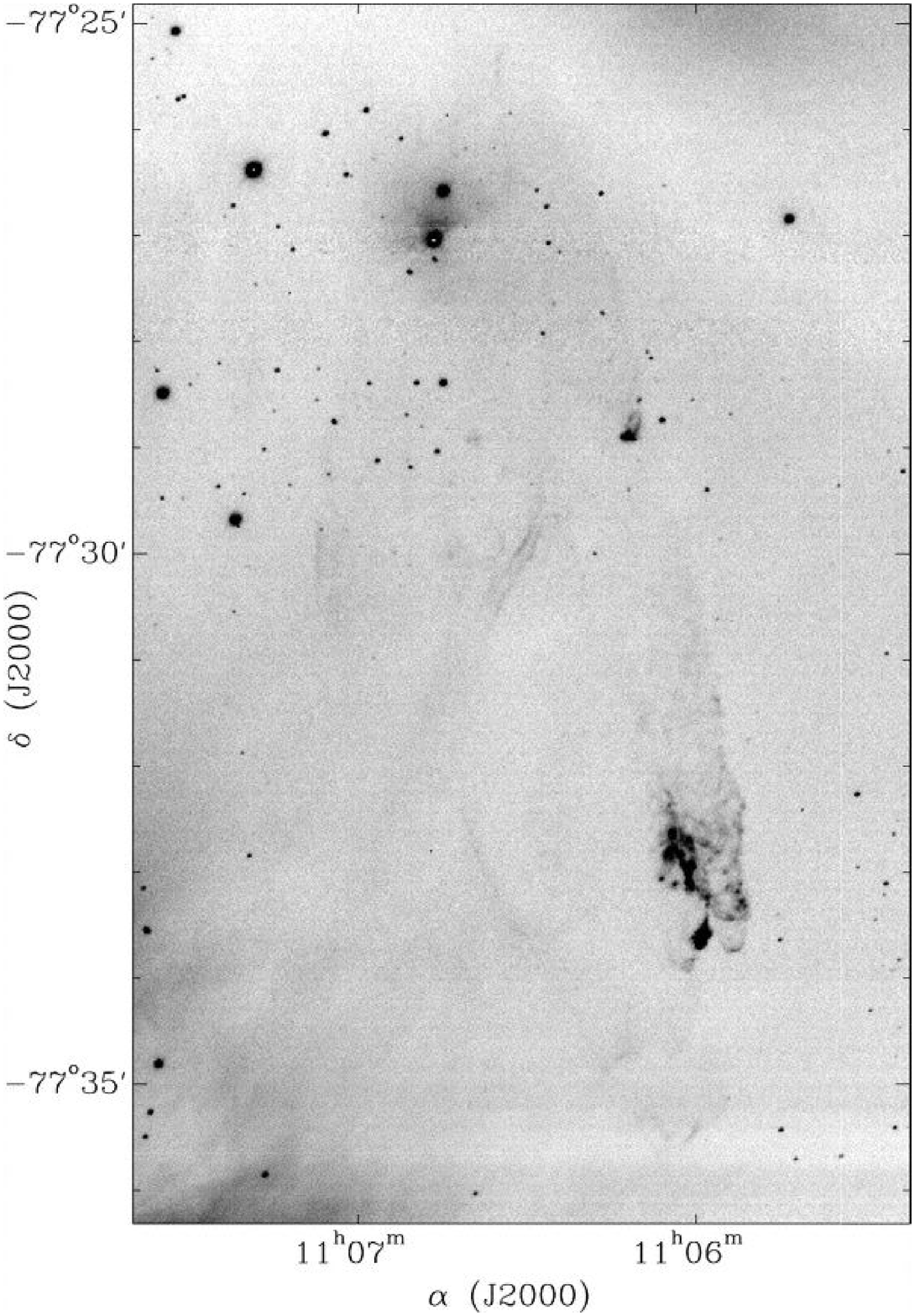}
\vspace{-2mm}
\caption{
H$\alpha$ image of HH~49, 50, 906, and 925 in Cha~I \citep{bal06}.
}
\label{fig:halpha}
\end{figure}

During the objective prism survey of southern star-forming regions
by \citet{sch77}, several Herbig-Haro (HH) nebulae were discovered in
Cha~I and II, consisting of HH~48-50 in the former and HH~52-54 in the latter.
\citet{sch84} measured proper motions for these objects, which are
relatively large given the proximity of the Chamael-eon clouds.
They suggested that HH~49 and 50 are probably excited by B35, an embedded
star 10$\arcmin$ northeast of HH~50 and just east of Ced~110.
During a survey at 1300~\micron, \citet{rei96} discovered a candidate
protostar, Cha-MMS1, that could be the driving source for HH~49-50.
The star exciting these HH objects is probably also responsible for
a compact CO outflow found by \citet{pru91}.
However, through CO(3-2) observations, \citet{bel06} and \citet{hir07} found
that the outflow is produced by Ced~110~IRS4 rather than Cha-MMS1.
Near- and mid-IR images of Ced~110~IRS4 are shown in
Figures~\ref{fig:image} and \ref{fig:irs4}. The 24~\micron\ image in
Figure~\ref{fig:image} detects Cha-MMS1 \citep{bel06,luh08}.
A weaker outflow has been detected toward Ced~110~IRS6 \citep{hir07}.

A second protostar in Cha~I detected at millimeter
wavelengths by \citet{rei96} is probably the source of another molecular
outflow from \citet{mat89}. This millimeter source, Cha-MMS2, likely
corresponds to the young star T44 based on the agreement of their coordinates.
\citet{per99} suggested that a different protostar, ISO~192, is the exciting
source for the CO outflow from \citet{mat89}. \citet{per07} detected
elongated nebulosity near ISO~192 in the direction of the outflow.
However, Cha-MMS2/T44 is closer than ISO~192 to the center of the outflow.

\citet{wan06} performed a deep survey in [S~II] across a
$1\hbox{$^\circ$}\times2\hbox{$^\circ$}$ field encompassing the entire
Cha~I cloud, detecting a total of 18 HH objects.
\citet{bal06} discovered additional fainter HH objects in Cha~I
through imaging of most of the cloud in H$\alpha$, [S~II], $i\arcmin$, and
broad-band mid-IR filters with {\it Spitzer}. Figure~\ref{fig:halpha}
shows an H$\alpha$ image from \citet{bal06} for HH objects in Cha~I.
These surveys have found no clear evidence for outflows from brown dwarfs
in Cha~I. Through images in [S~II], \citet{com06} discovered an HH jet
from ESO~H$\alpha$~574. The faint photometric measurements of this object
suggest that it has a substellar mass, but \citet{luh07cha} classified
it as K7-M0, which is indicative of a solar-mass star. Thus, the faint
apparent luminosity of this object is probably due to an edge-on disk
\citep{com04}.

In Cha~II, \citet{gh88} obtained optical images and
spectra for HH~52-54 and measured their velocities.
Through millimeter CO imaging, \citet{knee92} discovered a pair of outflows
associated with HH~52-54. Possible exciting stars for HH~52-54 have
been discussed by \citet{gh88}, \citet{hug91}, and \citet{knee92}.
HH~54 has been studied in detail with mid- and far-IR {\it ISO} and
{\it Spitzer} spectroscopy of emission in CO, H$_2$O, H$_2$, [O I], [C II],
[Ne II], [Fe II], [S I], and [Si II] \citep{lis96,nis96,neu98,gia06,neu06}.

\section{Circumstellar Disks}

As with other aspects of star formation, Chamaeleon has been
a valuable laboratory for studies of circumstellar disks.
As discussed in Sections~\ref{sec:optnir} and \ref{sec:mirfir},
IR photometry has been used to find new members by searching for
objects that exhibit excess emission indicative of disks.
Data of this kind also have been obtained for known members to determine
if disks are present \citep{gla79}.
\citet{kg01} used $JHKL$ photometry to estimate the fraction of members
of Cha~I with disks, and \citet{jay03} extended these measurements
to lower masses with the Cha~H$\alpha$ objects.
Photometry at longer wavelengths has been measured for most of the known
members of Cha~I and II through wide-field imaging surveys with the
{\it Spitzer Space Telescope}
\citep{you05,all06,por07,alc08,dam07,luh05ots,luh05frac,luh05disk2,luh08,luh08cha2}.
Pointed observations with {\it Spitzer} also have been performed toward
young stars scattered across Chamaeleon that were found by \citet{cov97a}
with {\it ROSAT} \citep{pad06}.
In these {\it Spitzer} data, mid-IR excess emission was detected for the
faintest known members of Cha~I, making them the least massive brown
dwarfs observed to have circumstellar disks
\citep{luh05ots,luh05disk2,luh08,luh08cha2}.
\citet{luh05frac} measured disk fractions for 109 low-mass stars and
brown dwarfs in Cha~I and presented 3.6-8~\micron\ photometry for the
objects later than M6. Using a subset of the images from \citet{luh05frac},
\citet{dam07} performed a separate measurement of photometry for 81 known
members and repeated the disk fraction measurement from \citet{luh05frac}.
\citet{luh08} and \citet{luh08cha2} analyzed all {\it Spitzer} images at
3.6-24~\micron\ that
have been obtained near Cha~I and presented compilations of the photometry
for more than 200 known members appearing within those data.
\citet{luh08} found that their photometric errors were much smaller than
those reported by \citet{dam07} for the same objects measured from the
same images.

\begin{figure}[!ht]
\plottwo{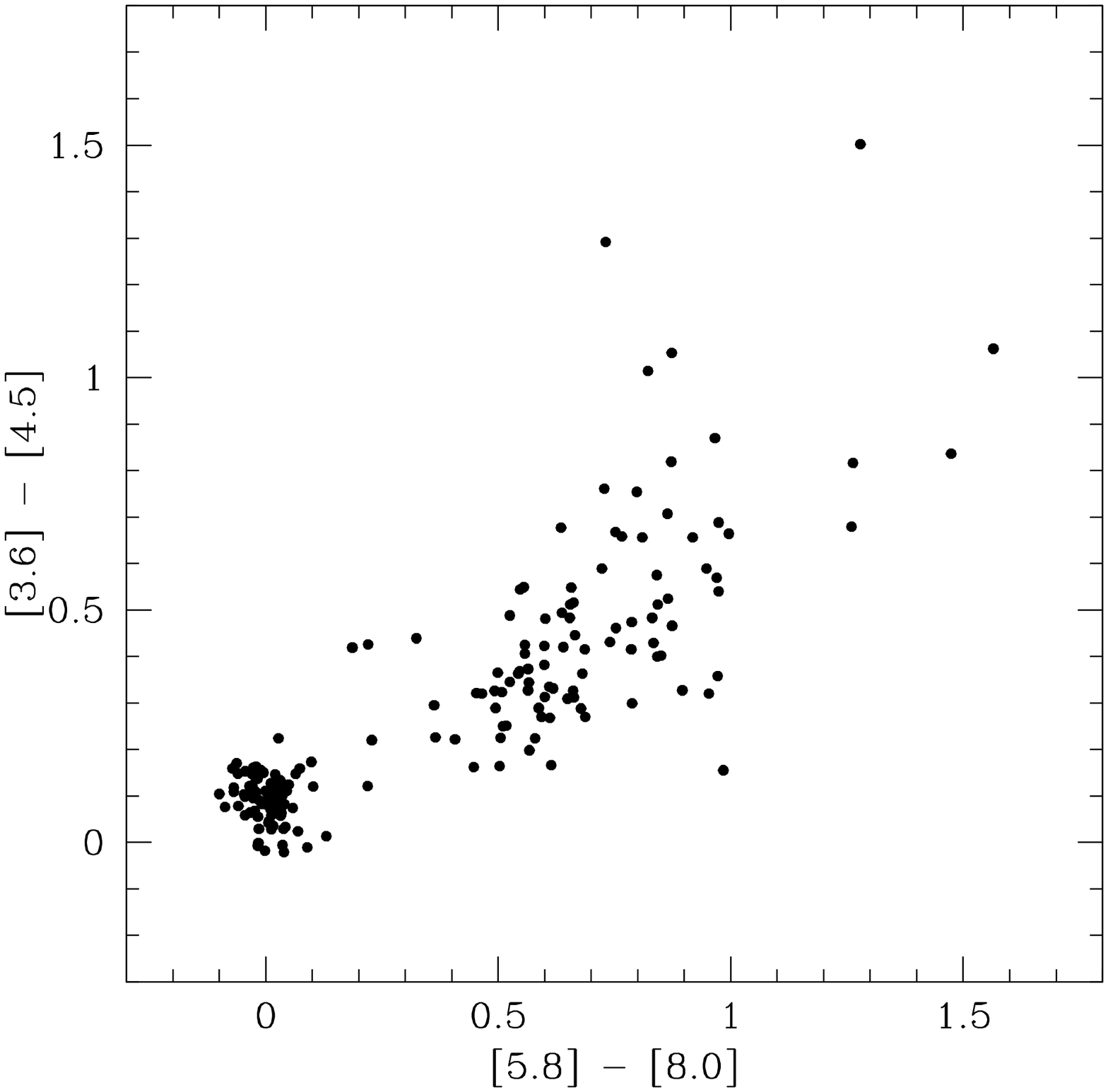}{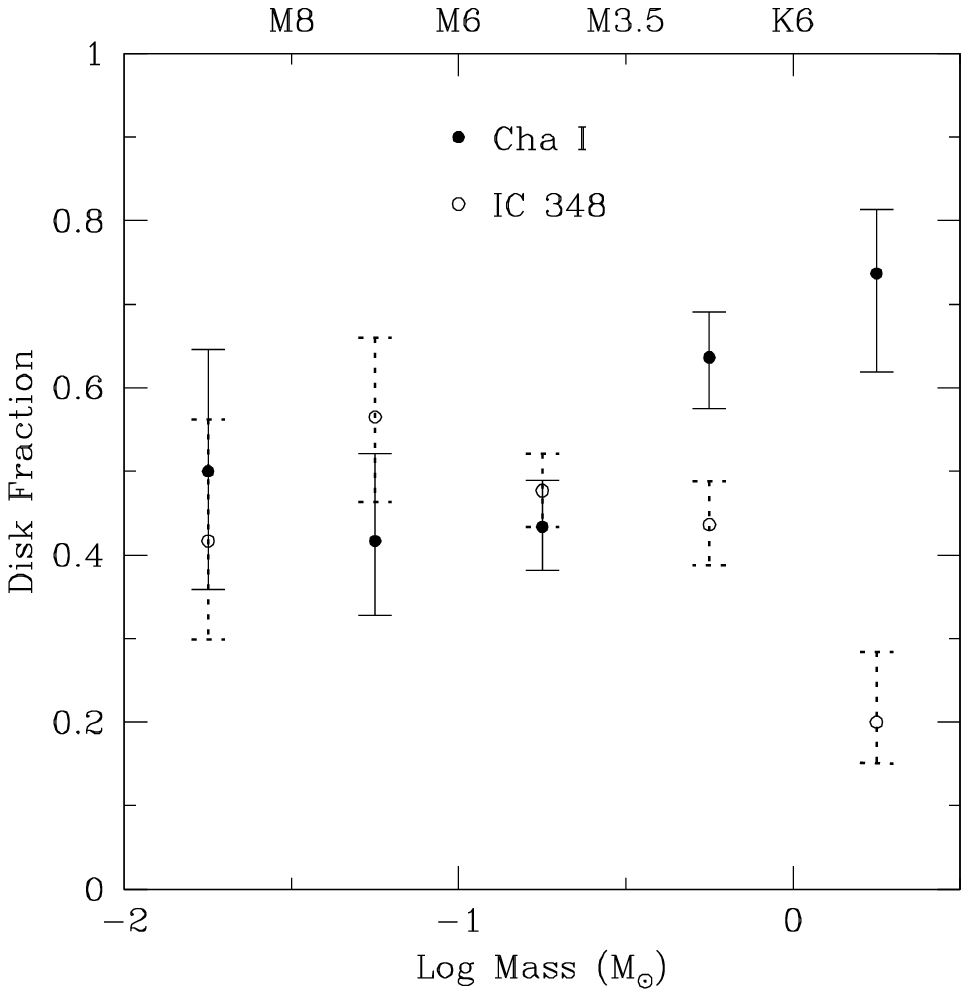}
\caption{
{\it Left}: Mid-IR color-color diagram for known members of Cha~I
\citep{luh08,luh08cha2}.
Objects with inner disks and envelopes exhibit red colors that
are distinctive from the neutral, photospheric colors of diskless members.
{\it Right}: Disk fractions as a function of mass and spectral type for
members of Cha~I \citep{luh08} and IC~348 \citep{luh05frac,lada06}.}
\label{fig:spitzer}
\end{figure}

By combining the {\it Spitzer} photometry with the extensive census
of known members that is available for Cha~I, \citet{luh08}
investigated several aspects of the disk population in this cluster.
They used the {\it Spitzer} colors to identify the members with
disks, as illustrated in Figure~\ref{fig:spitzer}, and
measured the fraction of sources with disks as a function of mass
from 0.01 to 3~$M_\odot$. As shown in Figure~\ref{fig:spitzer},
the disk fraction for solar-type stars is higher in Cha~I than in IC~348,
even though the two clusters have the same age.
\citet{luh08} suggested that the lifetimes of disks around solar-mass stars
in Cha~I may be longer because of the lower stellar density.
In addition, because many of the members of Cha~I were observed at multiple
epochs with {\it Spitzer}, \citet{luh08} and \citet{luh08cha2}
were able to demonstrate that stars
with disks have higher levels of mid-IR variability than diskless stars.
\citet{alc08} performed a similar analysis of the disk population in
Cha~II by combining the {\it Spitzer} photometry from \citet{you05} and
\citet{por07} with data at other wavelengths.

Disk-bearing members of Cha~I and II identified in photometric surveys
have been examined in detail through optical diagnostics of accretion
and mid-IR spectroscopy and spectral energy distributions.
For large samples of low-mass stars and brown dwarfs in Cha~I,
\citet{nat04}, \citet{muz05}, and \citet{moh05} searched for evidence of
accretion through high-resolution spectroscopy of emission in
H$\alpha$, Ca~II, and near-IR hydrogen transitions.
\citet{nat01}, \citet{apa02}, and \citet{ster04} constructed mid-IR spectral
energy distributions for the low-mass objects Cha~H$\alpha$~1, 2,
and compared the results to predictions of disk models.
\citet{bar08} detected rovibrational H$_2$ emission from
six members of Cha~I (T11, T2, T32, T33A, T33B, T43), which they
attribute to gas in circumstellar disks.
Through mid-IR spectroscopy with {\it ISO} and {\it Spitzer},
grain growth and crystallinity have been studied in disks
around solar-mass stars \citep{gur99,nat00,kes06} and low-mass stars
and brown dwarfs \citep{apa05} in Cha~I and Cha~II.
Mid-IR spectroscopy with {\it Spitzer} also has been used to characterize
a disk with an inner hole, known as a transitional disk,
around the solar-mass star CS~Cha \citep{esp07}.
The presence of an inner hole in this disk was originally suggested by
observations of jet emission \citep{tak03} and is probably at least partially
caused by a stellar companion \citep{gue07}.
Additional candidate transitional disks in Cha~I
have been identified through {\it Spitzer} photometry
\citep{dam07,luh08,luh08cha2}.

Because of the close proximity of Cha~I, it has been possible to resolve
a few disks in this cluster through direct imaging.
For instance, the disk around the B star HD~97048 was imaged
through mid-IR emission from polycyclic aromatic hydrocarbons
\citep{lag06,dou07} and optical scattered light images obtained with
the {\it Hubble Space Telescope} \citep{doe07}.
{\it Hubble} imaging also has resolved an edge-on disk around a
low-mass star in Cha~I \citep{luh07cha}.

\acknowledgments
This work was supported by grant AST-0544588 from the National Science
Foundation. I thank Gerald Rhemann and Hans Zinnecker for providing the images
for Figures~\ref{fig:map3}, \ref{fig:irn}, and \ref{fig:irs4}.
This publication has made use of NASA's Astrophysics Data System
Bibliographic Services, the SIMBAD database, operated at CDS in Strasbourg,
France, and data products from 2MASS and DSS.
2MASS is a joint project of the University of Massachusetts
and the Infrared Processing and Analysis Center/California Institute
of Technology, funded by the National Aeronautics and Space
Administration and the National Science Foundation.
DSS was produced at the Space Telescope Science Institute under
U.S. Government grant NAG W-2166.

\end{document}